\documentclass[12pt]{article}
\usepackage{jheppub}

\makeatletter
\def\@fpheader{\relax}
\makeatother

\usepackage{amssymb,amsmath,amsfonts,stmaryrd,mathscinet}
\usepackage{xcolor}

\usepackage{tikz}
\usetikzlibrary{decorations.markings,arrows,decorations.pathreplacing}

\newcommand\arXiv[1]{\href{http://arxiv.org/abs/#1}{\texttt{arXiv:#1}}}
\newcommand\MRnumber[2]{}
\newcommand\DOI[1]{\href{http://dx.doi.org/#1}{\texttt{DOI:#1}}}

\newtheorem{puzzle}{Puzzle}

\usepackage{dsfont}
\renewcommand\mathbb\mathds

\newcommand\bC{\mathbb C}

\newcommand\bH{\mathbb H}

\newcommand\bQ{\mathbb Q}
\newcommand\bR{\mathbb R}

\newcommand\bZ{\mathbb Z}

\newcommand\cB{\mathcal B}

\newcommand\cF{\mathcal F}

\newcommand\cH{\mathcal H}

\newcommand\cN{\mathcal N}

\newcommand\cX{\mathcal X}
\newcommand\cY{\mathcal Y}

\usepackage[mathscr]{euscript}

\renewcommand\d{\mathrm d}

\newcommand\MF{\mathrm{MF}}

\newcommand\longto\longrightarrow
\newcommand\mono\hookrightarrow
\newcommand\epi\twoheadrightarrow

\newcommand\<\langle
\renewcommand\>\rangle
\newcommand\sminus\smallsetminus

\DeclareMathOperator\Fer{Fer}
\DeclareMathOperator\Cliff{Cliff}
\DeclareMathOperator\Loops{Loops}

\newcommand\define[1]{\emph{#1}}

\title{Mock modularity and a secondary elliptic genus}
\author{Davide Gaiotto and Theo Johnson-Freyd}
\affiliation{Perimeter Institute for Theoretical Physics \\ Waterloo, ON, CANADA}
\emailAdd{dgaiotto@perimeterinstitute.ca}\emailAdd{tjohnsonfreyd@perimeterinstitute.ca}
\abstract{The theory of Topological Modular Forms suggests the existence of deformation invariants for two-dimensional supersymmetric field theories
that are more refined than the standard elliptic genus. In this note we give a physical definition of some of these invariants. The theory of mock modular forms makes a 
surprise appearance, shedding light on the integrality properties of some well-known examples. }
\preprint{}
\date{\today}

\begin{document}

\maketitle

\section{Introduction}

In \cite{WittenTMF} the following puzzle was posed; the goal of this paper is to propose a solution. Let us say that a (1+1)-dimensional quantum field theory with minimal, aka $\cN=(0,1)$, supersymmetry is \define{null} if supersymmetry is spontaneously broken and \define{nullhomotopic} if it can be connected to a null theory by a sequence of deformations, including deformations that may zig-zag up and down along RG flow lines.

\begin{puzzle} \label{mainpuzzle} Show that the supersymmetric quantum field theory of three free antichiral fermions $\bar\psi_1,\bar\psi_2,\bar\psi_3$ and supersymmetry generated by $G = \sqrt{-1}\, {:}\bar\psi_1\bar\psi_2\bar\psi_3{:}$ is not nullhomotopic. \end{puzzle}

For the remainder of this paper we will write simply ``SQFT'' for ``(1+1)-dimensional quantum field theory with $\cN=(0,1)$ supersymmetry'', and ``SCFT'' for an SQFT which is furthermore superconformal.
 The SQFT in Puzzle~\ref{mainpuzzle} is an SCFT, and is the (conjectured) limit under RG flow of the $\cN=(0,1)$ sigma model with target the round $S^3$ and minimal nonzero B-field. The puzzle is difficult because, as is shown in \cite{WittenTMF}, the direct sum of 24 copies of the SQFT in  Puzzle~\ref{mainpuzzle} \emph{is} nullhomotopic (as is the $\cN=(0,1)$ sigma model with target $S^3$ and B-field of strength $24$). So the puzzle requires constructing a torsion-valued deformation-invariant of SQFTs that is more sensitive than the 
  elliptic genus.

The motivation for the puzzle comes from the theory of Topological Modular Forms (TMF) described for example in \cite{MR1989190,MR3223024}. Based on suggestions from \cite{MR992209}, it is conjectured in \cite{MR2079378,MR2742432} that every SQFT defines a class in TMF, invariant under deformations of the SQFT. Indeed, \cite{MR2079378,MR2742432} conjecture that this TMF class exactly captures the deformation class of the corresponding SQFT. The TMF-valued invariant of an SQFT refines the usual modular-form valued elliptic index.  It is known that the TMF-valued invariant of $S^3$ (with minimal nonzero B-field) has exact order~$24$, hence the puzzle.

In this paper we will propose a solution to this puzzle. Let $\cB$ be an SCFT which, like the one in Puzzle~\ref{mainpuzzle}, has gravitational anomaly $c_R - c_L = 3/2$.%
\footnote{Our invariant applies to SQFTs of gravitational anomaly $c_R - c_L \in -\frac12 + 2\bZ$.} 
If $\cB$ is not initially conformal, flow it to the IR before proceeding. Build:
\begin{enumerate}
\item a ``generalized mock modular form $f_1$'' with source equal to the torus one-point function of the supersymmetry generator of $\cB$.%
\footnote{Our convention will be to use powers of $\eta$ to correct for multipliers in (mock) modular forms. The ``generalized mock modularity'' condition we demand for $f_1$ is that $f_1(\tau) = \lim_{\bar\tau \to -i\infty} \hat{f}_1(\tau,\bar\tau)$ where $\hat{f}_1(\tau,\bar\tau)$ is modular invariant with appropriate weights and
$$ \sqrt{-8\tau_2} \partial_{\bar\tau} \hat{f}_1(\tau,\bar\tau) = \eta(\tau)^3 \times (\text{one-point function of } \bar{G}).$$
As usual, $\tau_2 = \frac1{2i}(\tau - \bar\tau)$. One must make various essentially-arbitrary sign choices, one of which is the choice of square root $\sqrt{-8}$.}
 The $q$-expansion of $f_1$ is not determined by $\cF$, but the class of $[f_1] \in \bC(\!(q)\!) / \MF_2$ is.
\item a nonnegative-integral $q$-series $f_2$ equal to half the graded dimension of the space of {bosonic} Ramond-sector ground states in $\cB$.%
\footnote{Because of the gravitational anomaly $c_R - c_L = \frac32 = -\frac12 + 2 \times (\text{odd})$, the vector space of bosonic ground states is automatically pseudoreal aka quaternionic, and so of even complex dimension.  If $c_R - c_L$ were instead $-\frac12 + 2 \times (\text{even})$, then the bosonic ground states would be a real vector space, $f_2$ would take half-integral values, and we would consider its class mod $\bZ(\!(q)\!)$ rather than mod $2\bZ(\!(q)\!)$.}
The class $[f_2] \in \bC(\!(q)\!) / 2\bZ(\!(q)\!)$ is a sort of ``mod-2 index'' of~$\cF$.
\end{enumerate}
Neither class $[f_1] \in \bC(\!(q)\!) / \MF_2$ nor $[f_2] \in  \bC(\!(q)\!) / 2\bZ(\!(q)\!)$ is a deformation invariant of~$\cB$. But we will argue that the class $[f_1] - [f_2] \in \bC(\!(q)\!) / [ \MF_2 + 2\bZ(\!(q)\!)]$ is invariant under SQFT deformations. Furthermore, we will compute that for the SQFT in Puzzle~\ref{mainpuzzle}, this invariant is nonzero and in fact has exact order $24$ in $\bC(\!(q)\!) / [ \MF_2 + 2\bZ(\!(q)\!)]$. This is our solution to Puzzle~\ref{mainpuzzle}.

Homotopy-theoretic considerations imply the existence (and nontriviality) of an invariant like ours \cite{DBE2015}, which when restricted to sigma models is described both analytically and geometrically in \cite{MR3278648} (where it is also shown that the invariant of $S^3$ has exact order $24$). But topological arguments do not explain how to compute this invariant of SQFTs except when the SQFT can be deformed to a sigma model. Our description of the invariant connects it explicitly to holomorphic anomalies of noncompact SQFTs, and thereby to mock modularity, which is of great current interest due to the ``moonshine'' of \cite{MR2802725,MR3271175}.

Whenever the CFT $\cB$ is rational, the source of the holomorphic anomaly equation is a modular-invariant bilinear combination of vector-valued holomorphic and anti-holomorphic modular forms. 
The corresponding generalized mock modular form is then 
a ``mixed mock modular form'':
a bilinear combination of the same vector-valued holomorphic modular form and 
a true vector-valued mock modular form. In such a situation, our arguments can be seen as a justification for the 
very existence of mock modular forms with interesting integrality properties.

The paper is structured as follows. 

Section~\ref{sec.general} presents the general story. 
We first make some brief remarks about gravitational anomalies, and a $\eta(\tau)^n$ normalization factor that we include in the elliptic genus\footnote{We will use the term ``Witten genus'' for the combination $\eta(\tau)^n\times(\text{elliptic genus})$.}
 and in one-point functions in order to correct the multipliers that would otherwise be present.
 We then discuss the properties of 
certain SQFTs which violate the compactness constraint  in a controllable manner. It turns out to be still possible to define the elliptic genus of such SQFTs \cite{Eguchi:2006tu,MR2821103,Ashok:2013pya,Gupta:2017bcp}.
Such elliptic genus satisfies a ``holomorphic anomaly'' equation with a source which we characterize in a precise manner 
in terms of the torus one-point function of the supersymmetry generator in a compact ``boundary SQFT''. We use this construction to argue that the torus one-point function of the supersymmetry generator 
in a nullhomotopic SQFT is the source of a holomorphic anomaly equation for a generalized mock modular form. We then argue that the coefficients of this generalized mock modular form are (even) integral, up to a correction arising as a type of ``mod 2 index'' of the boundary SQFT.
It follows that, if solutions of the holomorphic anomaly 
equation fail to have integral (plus correction) mock modular parts, the SQFT cannot be nullhomotopic. This is the justification for our invariants.
 We then recast our invariant in homotopy-theoretic terms, where it becomes the ``secondary invariant'' of \cite{MR3278648}. 

To illustrate our proposed invariant, we focus
  on two families of examples. First, in Section~\ref{sec.cigar}, we study the sigma models with target $S^1$ and the ``cigar.''
 We start by reviewing the $S^1$ sigma models, with an emphasis on the role that the target-space spin structure plays on the behaviour of the model. This provides a chance to illustrate the mod-2 index, and allows us to compute our invariant for the $T^3$ sigma model with its Lie group framing. We then analyze the ``cigar,'' which is a noncompact manifold with ``boundary'' $S^1$, and demonstrate explicitly that the corresponding sigma model enjoys our predicted holomorphic anomaly and integrality.

The second set of examples, which we study in Section~\ref{sec.S3}, are the ones from \cite{WittenTMF}: the $\cN=(0,1)$ sigma model with target the round $S^3$ and with B-field of strength $k$.
We first warm up with the $k=1$ case of Puzzle~\ref{mainpuzzle}, and then study the general case. In all cases, we find that our invariant is precisely $k \pmod {24}$, showing that the $S^3_k$ sigma model is nullhomotopic if and only if $k = 0 \pmod {24}$.
We then mention a few related constructions and puzzles: we build an antiholomorphic SCFT of central charge $c_R = 27/2$ which we expect to represent the 3-torsion element in $\pi_{27}\mathrm{TMF}$; and we speculate that $S^3_k$ is ``flavoured-nullhomoptic'' for all $k$, with the nullhomotopy given by a certain ``trumpet'' geometry with $\cN=(0,4)$ supersymmetry.

Let us end this introduction by emphasizing that we expect our invariant captures only some of the torsion in the space of $\cN=(0,1)$ SQFTs.%
\footnote{The analogous invariant of TMF classes captures all of the $3$-torsion in degree $n=-1 \pmod 4$, but only some of the $2$-torsion \cite{MR3278648}.}
It is known that the TMF classes represented by the group manifolds (with their Lie group framings)
$$\mathrm{Sp}(2),\qquad G_2,\qquad G_2\times U(1)$$ are nonzero: their exact orders are, respectively, $$3, \qquad 2, \qquad 2.$$ 
The same logic as in \cite{WittenTMF} suggests that the $\cN=(0,1)$ sigma models for $\mathrm{Sp}(2)$ and $G_2$ flow in the IR to SCFTs consisting purely of ($10$ and $14$, respectively) antichiral free fermions, with supersymmetries that encode the structure constants of the Lie algebras $\mathfrak{sp}(2)$ and $\mathfrak{g}_2$.%
\footnote{The connection between free fermion SCFTs and Lie algebras is described in \cite{MR791865}. The SCFT in Puzzle~\ref{mainpuzzle} corresponds to the Lie algebra $\mathfrak{su}(2)$.}
The sigma model with target $G_2\times U(1)$ does not flow to a purely-antichiral theory, but rather to the product of the $\mathfrak{g}_2$-theory and the ``standard'' circle theory studied in \S\ref{subsubsec.S1-nonbounding}.
The elliptic and mod-2 indexes of all three SQFTs vanish. Moreover, the invariant described in this paper vanishes for all three SQFTs --- the first two for degree reasons, but the third nontrivially.
Due to the expected relationship between TMF and SQFTs, we expect that these SQFTs are not nullhomotopic. We leave the reader with the following puzzles:
\begin{puzzle}\label{sp2puzzle}
  \begin{enumerate}
  \item Show that the SQFT $\overline{\Fer}(\mathfrak{sp}(2))$ of $10$  antichiral free fermions and supersymmetry encoding the structure constants of $\mathfrak{sp}(2)$ is not nullhomotopic. 
  \item Show that the SQFT $\overline{\Fer}(\mathfrak{g}_2)$ of $14$ antichiral free fermions and supersymmetry encoding the structure constants of $\mathfrak{g}_2$ is not nullhomotopic. 
  \item Show that the product of $\overline{\Fer}(\mathfrak{g}_2)$ with a (standard) $S^1$ sigma model is not nullhomotopic.
  \end{enumerate}
\end{puzzle}

\section{A torsion invariant of SQFTs} \label{sec.general}

\subsection{Gravitational anomalies and spectator fermions}\label{sec.anomaly}

For a quantum field theory to be \define{gravitationally nonanomalous}, its partition function must be valued in numbers (as opposed to a section of some line bundle on the moduli space of spacetimes); it must have a well-defined Hilbert space (as opposed to a section of some gerbe); and so on for higher-codimensional data \cite{FreedTeleman2012}. In the $(1+1)$-dimensional SQFT case, the  \define{elliptic genus} $Z_{RR}$ is by definition the partition function on flat tori with nonbounding, aka Ramond--Ramond aka periodic--periodic, spin structures. The moduli space of Ramond--Ramond flat tori is three-real-dimensional --- in addition to the complex and anticomplex parameters $(\tau,\bar\tau)$, there is also a ``size'' parameter --- but we will generally compute in the IR aka large-torus limit. In this limit, the partition function of a nonanomalous $(1+1)$-dimensional SQFT will definitely be modular for the full $\mathrm{SL}(2,\bZ)$ with weight $(0,0)$ and no multiplier.%
\footnote{Real-analytic modular forms have two weights. A weight $(w,w')$ modular form transforms under $\bigl( \begin{smallmatrix} a & b \\ c & d \end{smallmatrix}\bigr) \in \mathrm{SL}(2,\bZ)$ with a factor of $(c\tau+d)^w(c\bar\tau+d)^{w'}$.} 
Indeed, modularity is transparent from the path-integral description.%
\footnote{A priori, the partition function can blow up at the cusp in a manner controlled by the IR central charges of the theory. If the SQFT is ``compact'' then its index will be weakly holomorphic by a standard argument.}

The SQFT in Puzzle~\ref{mainpuzzle},
and more generally any $\cN=(0,1)$ sigma model, suffers a gravitational anomaly due to the unpaired antichiral fermions. For an SCFT, the gravitational anomaly is the difference between the left- and right-moving central charges; for an SQFT, these separate central charges are not well-defined, but the total gravitational anomaly is, and is preserved under RG flow.
We will normalize the gravitational anomaly so that for an SCFT with left and right moving central charges $c_L$ and $c_R$, the anomaly is $n := 2(c_R - c_L)$. The factor of $2$ is natural because then $n$ ranges over $\bZ$. The gravitational anomaly $n \in \bZ$ plays the role of homotopical degree in \S\ref{sec.BN}, and so we will occasionally refer to it as the \define{degree} of the SQFT.

The gravitational anomaly manifests in various ways. First of all, it produces a nontrivial multiplier for the behaviour of the elliptic genus under the $\tau \mapsto \tau+1$, namely $Z_{RR}(\tau+1,\bar\tau+1) = e^{-2\pi i n/24} Z_{RR}(\tau,\bar\tau)$; the path integral description still guarantees that $Z_{RR}$ transforms under $\tau \mapsto -1/\tau$ with weight $(0,0)$ and some multiplier.
Second, the gravitational anomaly leads to an ambiguity in the parity of ``the'' Ramond sector of the theory. This 
leads to a sign ambiguity when trying to define ``the'' elliptic index.

Our convention, standard in algebraic topology, will be to trade nontrivial multipliers for nontrivial weights of modular forms, by including a normalization factor of $\eta(\tau)^n$ whenever necessary, where $\eta(\tau) = q^{1/24} \prod_{j=1}^\infty (1-q^j)$ is Dedekind's eta function. The combination $\eta(\tau)^n Z_{RR}(\tau,\bar\tau)$ is sometimes called the \define{Witten genus} of a gravitationally-anomalous SQFT, and we will use that term. It is automatically modular without multiplier, of weight $(\frac n 2, 0)$.

When $n>0$, the Witten genus can be interpreted as follows. Consider the nonanomalous SQFT $\Fer(n) \otimes \cF$, where $\Fer(n) = \Fer(1)^{\otimes n}$ means the holomorphic CFT of $n$ chiral fermions $\psi_1,\dots,\psi_n$, acted upon trivially by the right-moving supersymmetry. Deformations of $\cF$ correspond to deformations of $\Fer(n) \otimes \cF$ which preserve the $\Fer(n)$-subsector. The $n$ free chiral fermions in $\Fer(n)$ are called \define{spectators}.
Because of the zero modes of the chiral fermions, the plain elliptic genus of $\Fer(n) \otimes \cF$ vanishes. But because we have a  distinguished $\Fer(n) \subset \Fer(n) \otimes \cF$, we find a distinguished observable, namely ${:}\psi_1\cdots\psi_n{:}$.
 The Witten genus of the gravitationally-anomalous SQFT $\cF$ is precisely the one-point function of $(-1)^{\frac{n}{4}} {:}\psi_1\cdots\psi_n{:}$ in $\Fer(n) \otimes \cF$.%
  \footnote{We include the $(-1)^{\frac{n}{4}}$ phase because the torus one-point function of ${:}\psi_1\psi_2{:}$ in $\Fer(2)$ equals $\sqrt{-1}\, \eta(\tau)^2$; compare equation~(\ref{eqn-ferbarfer}).} 
  By ``one-point function,'' we will always mean the torus one-point function on nonbounding, aka Ramond--Ramond, tori.

The spectator fermions furthermore allow the sign ambiguity to be handled by demanding that $\Fer(n) \otimes \cF$ have \emph{trivialized} anomaly, including a choice of parity for its Ramond-sector ground state.\footnote{When $n <0$, one cannot use spectator fermions to handle the sign ambiguity. One may instead call upon an equivalent discussion in terms of  relative quantum field theories in the sense of \cite{FreedTeleman2012}.}
More precisely, the sign ambiguity can be swapped for an ambiguity in the choice of generators of $\Fer(n)$. We will largely ignore the sign ambiguity in this paper, since we will not try to add different SQFTs together (and so we will never risk thinking we have found a cancellation when there was not one).

Another phenomenon in gravitationally-anomalous SQFTs becomes transparent when working with spectator fermions. Consider the Ramond-sector Hilbert space $\cH_R$ for the spectated theory $\Fer(n) \otimes \cF$. The decoupled $\Fer(n)$ subalgebra of the full observable algebra provides operators on this Hilbert space: specifically, $\cH_R$ is naturally a module for the fermion zero modes,%
\footnote{The Fourier expansion of a fermion in the Ramond sector is integrally-graded.}
 which form a copy of the $n$th Clifford algebra $\Cliff(n)$.
Moreover, the SQFT $\Fer(n) \otimes \cF$ compactified on $S^1$ automatically possesses a time-reversal structure --- showing this is an interesting exercise, solved in \cite[\S3.2.2]{GPPV} --- and hence $\cH_R$ possesses a real form, acted on by the  real Clifford algebra $\Cliff(n,\bR)$. Modules for $\Cliff(n,\bR)$ represent degree-$n$ classes in oriented K-theory.

\subsection{Non-compact SQFTs}\label{sec.noncompact}

The discussion in \S\ref{sec.anomaly} implicitly assumed that the SQFT in question was ``compact'' --- for instance, that its Hamiltonian had sufficiently discrete spectrum --- in order for its elliptic genus and K-theory class to be well-defined.
There are at least two distinct ways one can enlarge the space of SQFTs which admit some kind of elliptic genus. 

\subsubsection{Flavoured-compact theories}\label{subsec.flavoured-compact}
The first way is to consider SQFTs equipped with a continuous global symmetry, say $U(1)$ for simplicity, and define a ``flavoured'' elliptic genus 
as a partition function on a torus equipped with a flat $U(1)$ connection. The flat connection can be parameterized by a point $\xi$ in the elliptic curve $E_\tau$ of parameter $\tau$. 
As long as the current sits in a standard $(0,1)$ multiplet, there is a superpartner of the 
anti-holomorphic part of the current which should guarantee the independence of the partition function on the anti-holomorphic part $\bar \xi$ of the connection,
leaving a holomorphic dependence on $\xi$. As a consequence, the flavoured Witten genus is a Jacobi form, of weight $\frac{n}{2}$ and index $\ell$ determined by the 
't Hooft anomaly coefficient for the $U(1)$  global symmetry. 

Such a flavoured SQFT can be considered ``compact'' as long as the Hamiltonian on the circle has sufficiently discrete spectrum for a generic choice of flat $U(1)$ connection on the circle.
Then the Witten genus will be well-defined as a meromorphic Jacobi form. If the SQFT is compact even in the un-flavoured sense then the Witten genus will be a holomorphic Jacobi form and 
admit an expansion in terms of theta functions of index $\ell$, with coefficients which form a vector-valued modular form. 

For a sigma model, the calculation of the flavoured Witten genus will involve the equivariant analogues of the characteristic classes involved in the calculation of the standard Witten genus. 

The canonical example is a $(0,1)$ sigma model with target $\bR^2$ and a $U(1)$ isometry acting as rotations of $\bR^2$. The corresponding flavoured Witten genus 
 is a meromorphic \ Jacobi form of
 weight $1$ and index $-1$:

\begin{equation}\label{eqn.R2}
Z_{RR}(\bR^2)[\xi;\tau] = \frac{\eta(\tau)^3}{\theta(\xi;\tau)} =\frac{1}{x^{\frac12}-x^{-\frac12}}\prod_{n=1}^\infty \frac{(1-q^n)^2}{(1-x q^n)(1-x^{-1} q^n)}
\end{equation}
with $q = \exp 2 \pi i \tau$ and $x = \exp 2 \pi i \xi$.
The analogue of TMF for the flavoured SQFTs does not appear to be well-studied \cite{GPPV}. It would be very interesting to do so.  

\subsubsection{Theories with cylindrical ends}
The second way we can enlarge the space of SQFTs which admit some kind of elliptic genus is by considering the SQFT analogue of sigma models on manifolds 
with an asymptotic boundary region which approaches $\bR^+ \times B$ for some compact manifold~$B$. We can formalize this notion by requiring the SQFT $\cF$ 
to be equipped with a local operator $\Phi$ with the following property:
\begin{itemize}
\item Consider the direct product of $\cF$ and a free Fermi multiplet, i.e.\ a free chiral fermion~$\lambda$ annihilated by the supercharge.\footnote{In the sense of \S\ref{sec.noncompact}, the fermion $\lambda$ is the ``spectator'' corresponding to the noncompact direction.}
\item Deform the product theory by a ``fermionic superpotential'' $\lambda (\Phi-p)$, i.e.\ add the terms $\lambda (\bar G_0 \Phi) - (\Phi-p)^2$  
to the Lagrangian, where $p \in \bR$ is a parameter. 
\item The result is a family of SQFTs $\cB_p$ parameterized by a point $p$ in $\bR$. We require $\cB_p$ to stabilize to some compact SQFT $\cB$ for large positive $p$ 
and $\cB_p$ to spontaneously break supersymmetry for large negative $p$. 
\end{itemize}
Note that in particular the family $\cB_p$ built from $\cF$ is a nullhomotopy of the compact SQFT $\cB$. Conversely, any nullhomotopy of $\cB$ can be converted into a noncompact SQFT $\cF$ by reversing the steps above, i.e.\ promoting the parameter of the deformation family to a dynamical $(0,1)$ chiral multiplet. 

What is the elliptic genus of $\cF$? The question is subtle because, being noncompact, $\cF$ has continuous spectrum in its Hamiltonian.
 Because of this, the usual supersymmetric cancelation argument verifying that the elliptic genus is holomorphic breaks down. Let us work in the IR limit.
  In this limit, the Witten genus $\eta(\tau)^n Z_{RR}(\cF)(\tau,\bar\tau)$  will automatically transform as a weight $(\frac n2,0)$ modular form under the action of $\mathrm{SL}(2,\bZ)$ acting simultaneously on both $(\tau,\bar\tau)$, since modularity is manifest from the path-integral description of the index. 
  
  Let us try to prove that $Z_{RR}(\cF)$ is holomorphic. We will fail, and by failing we will instead compute the \define{holomorphic anomaly} of $\cF$.
We will try to apply the usual proof of holomorphicity of the index. For any QFT,%
\footnote{In the IR limit, the stress-energy tensor has only two terms, the chiral and antichiral parts $T(z),\bar{T}(\bar z)$. In the nonconformal case, the stress-energy tensor picks up a third component --- its ``trace'' --- measuring the dependence of $Z_{RR}$ on the size of the worldsheet torus.}
\begin{equation}
 \frac{\partial}{\partial\bar{\tau}}Z_{RR}(\cF) = - 2\pi i (\text{torus one-point function of } \bar{T}  \mbox{ in $\cF$}).
\end{equation}
Use the supersymmetry: $\bar{T} = \frac12 [\bar{G}_0,\bar{G}]$, where the commutator $[,]$ means the supercommutator, $\bar{G}_0$ is the generator of $\cN=(0,1)$ supersymmetry
and $\bar{G}$ the anti-holomorphic component of the supercurrent. In a compact theory, the one-point function of any anti-commutator $[\bar{G}_0, O]$ would vanish. 
In a non-compact theory, with non-compact direction parameterized by the expectation value of the operator $\Phi$, we can imagine integrating by parts in the space of fields to obtain 
a term proportional to%
\footnote{The precise proportionality factor is hard to derived from path integral considerations. Rather, the factor of $\sqrt{-8}$ in (\ref{eqn.holomorphicanomaly}) comes from careful computation of examples in Section~\ref{sec.cigar}.}
the torus one-point function of $O$ evaluated in the boundary theory $\cB$,
resulting in the following \define{holomorphic anomaly equation}.
 Including the spectator fermions to fix the normalizations, we propose:%
 \footnote{It would be interesting to work out the relation between this holomorphic anomaly equation and the 
holomorphic anomaly equation which occurs for {\it modified} elliptic genera of the world-volume theories of E-strings \cite{Minahan:1998vr}, MSW $M5$-strings \cite{Manschot:2007ha} or Vafa-Witten partition functions \cite{Manschot:2017xcr}. }
\begin{multline}\label{eqn.holomorphicanomaly}
\sqrt{-8\tau_2} \frac{\partial}{\partial\bar{\tau}} \bigl[\text{Witten genus of }\cF\bigr] \\
=(\text{torus one-point function of } (-1)^{\frac{n}{4}}{:}\psi_1\cdots\psi_{n-1}\bar{G}{:}  \mbox{ in $\Fer(n-1) \times \cB$})
\end{multline}
Here $\tau_2 = \frac1{2i}(\tau - \bar\tau)$ is the imaginary part of $\tau$. 
The sign of the square root $\sqrt{-8}$ is essentially arbitrary, and can be absorbed in the ambiguity in the sign of $\bar{G}$ or in the sign of the Ramond sector of $\cB$.
As a reality check, note that, in the IR limit, both sides are real-analytic modular of weight $(\frac{n-1}2, \frac32)$ with trivial multipliers.

Being a bit loose with the phase of the torus one-point function, we can write equation~(\ref{eqn.holomorphicanomaly}) as 
\begin{equation}\label{eqn.holomorphicanomaly2}
\sqrt{-8\tau_2} \, \eta(\tau) \frac{\partial}{\partial\bar{\tau}} Z_{RR}(\cF) = (\text{torus one-point function of } \bar{G}  \mbox{ in $\cB$}).
\end{equation}

A simple generalization of this construction is to require $\cB_p$ to stabilize to some compact SQFT $\cB_+$ for large positive $p$ and to another 
compact SQFT $\cB_-$ for large positive $p$ for large negative $p$. Then we expect, up to a phase factor:
\begin{multline}\label{eqn.bordismZRR}
\sqrt{-8\tau_2} \frac{\partial}{\partial\bar{\tau}}\bigl[\text{Witten genus of }\cF\bigr]  = \bigl[ (\text{torus one-point function of } {:}\psi_1\cdots\psi_n{:}\bar{G}  \mbox{ in $\cB_+$}) \\  -(\text{torus one-point function of } {:}\psi_1\cdots\psi_n{:}\bar{G}  \mbox{ in $\cB_-$}) \bigr]
\end{multline}
This means that, although the torus one-point function of $\bar{G}$ is not a deformation-invariant of an SQFT, it changes in a controlled fashion. This control is the basis of our torsion invariant.

\subsection{Integrality of the $q$-expansion} \label{sec.int}

In addition to holomorphicity, the other fundamental fact about the Witten genus $\eta^n Z_{RR}(\cF)$ of a (compact) SQFT $\cF$ is the integrality of its $q$-expansion. We briefly review the argument. Let $\cF[S^1]$ denote the $S^1$-equivariant supersymmetric quantum mechanics model produced by compactifying $\cF$ on a circle (with Ramond aka nonbounding spin structure). Then $\eta^n Z_{RR}(\cF)$ can be interpreted as the supersymmetric index of $\cF[S^1]$, with $q$ parameterizing the $S^1$-action, and indices are well-known to be integral, since they merely count with signs the number of supersymmetric ground states. Note that the compactification breaks manifest modularity. Indeed, suppose we didn't already know that $Z_{RR}(\cF)$ was holomorphic (for $\cF$ compact). Then this compactification implements the canonical way to extract a holomorphic function from a real-analytic modular form: analytically continue away from $\bar \tau = \tau^*$ and 
take the limit $\bar \tau \to -i \infty$.%
\footnote{The elliptic curve $E_\tau = \bC / (\bZ \oplus \tau\bZ)$ living in the $\bar\tau \to -i\infty$ limit is called the \define{Tate curve}.
It does not correspond to a Euclidean 2-torus because we explicitly broke the relation between $\tau$ and~$\bar\tau$. When $\tau$ is pure-imaginary, the Tate curve may be pictured as a nonrelativistic torus in which the ``space'' and ``time'' axes point in the $z$ and $\bar z$ directions, and the ``space'' direction is infinitely small compared to the ``time'' direction.
}

Depending on the value of the gravitational anomaly $n$, one can make stronger statements than mere integrality. The $q$-dependence is immaterial --- the statements hold in general for SQM models of degree $n$ equipped with a time-reversal symmetry.%
\footnote{The time-reversal structure on $\cF[S^1]$ arises from $180^\circ$-rotation of $\cF$ through a slightly subtle argument \cite[\S3.2.2]{GPPV}.}
 There are various ways to define the notion of ``degree-$n$ SQM model'', just like the choices in \S\ref{sec.anomaly} for how to handle the gravitational anomaly.
 One general way is to work with SQM models that are relative, in the sense of \cite{FreedTeleman2012}, to certain short-range-entangled $(1+1)$-dimensional phases.
 When $n\geq0$, another explicit method is to employ spectator fermions. Then a \define{degree-$n$ SQM model} is an SQM model (i.e.\ a super Hilbert space $\cH$ with an odd operator $G$ generating the supersymmetry; it is ``compact'' when $G$ is Fredholm) equipped with an action by the $n$th Clifford algebra $\Cliff(n)$ (which should (super)commute with $G$).
The presence of a time-reversal symmetry equips $\cH$ with a real structure, acted on by the real Clifford algebra $\Cliff(n,\bR)$. The supersymmetric ground states are then a finite-dimensional $\Cliff(n,\bR)$-module $V$.%
\footnote{The standard convention is that $\Cliff(n,\bR)$ is the real superalgebra with $n$ odd generators $\gamma_1,\dots,\gamma_n$, anticommuting with each other and each squaring to $-1$. We will later use the reasonably-standard notation $\Cliff(-n,\bR)$ to mean the algebra with the same odd generators but with $\gamma_i^2 = +1$. But in fact these conventions are purely arbitrary: one could decide, with no change in the final results, that $\gamma_i^2 = +1$ in $\Cliff(n,\bR)$ and that $\gamma_i^2 = -1$ in $\Cliff(-n,\bR)$. The reason for this arbitrariness is that the category of real superalgebras admits an automorphism exchanging $\Cliff(n,\bR)$ with $\Cliff(-n,\bR)$. As with all sign conventions, what is important is to be consistent.}

The usual {supersymmetric index} of the SQM model ignores the time-reversal symmetry: it depends just on $V\otimes \bC$ as a $\Cliff(n,\bC)$-module. When $n$ is even, $\Cliff(n,\bC)$ has two irreducible modules, differing by parity. Choose one of them arbitrarily to be ``the'' irrep $I$; then $V \otimes \bC \cong I^{a|b} = I \otimes_\bC \bC^{a|b}$, where $\bC^{a|b}$ means the (complex) supervector space with graded dimension $(a,b)$.  The ordinary \define{index} of $V$ is simply $a-b$.%
\footnote{Note that we again encounter an ambiguity in the sign of the index.}
But in the presence of a time-reversal symmetry, we don't just have the $\Cliff(n,\bC)$-module $V \otimes \bC$ --- we have the $\Cliff(n,\bR)$-module $V$. It turns out that when $n = 2 \pmod 4$, there is only one irreducible $\Cliff(n,\bR)$-module $J$, with complexification $J \otimes \bC \cong I^{1|1}$. Thus the index  vanishes when $n = 2 \pmod 4$. And when $n = 4 \pmod 8$, there are two irreducible $\Cliff(n,\bR)$-modules, $J^{1|0}$ and $J^{0|1}$, but $J^{1|0} \otimes \bC = I^{2|0}$ splits as two copies of the irreducible $\Cliff(n,\bC)$-module, and so the index is automatically even.
In summary, the index of a degree-$n$ SQM model with time reversal symmetry lives in $m\bZ$ with:
\begin{equation}\label{eqn.mofn}
 m = \begin{cases} 
 1, & n = 0 \pmod 8, \\
 2, & n = 4 \pmod 8, \\
 0, & \text{else}.
\end{cases}
\end{equation}
In the SQFT case that we care about, the Witten genus $\eta^n Z_{RR}(\cF)$ has $q$-expansion in $m\bZ(\!(q)\!)$.%
\footnote{We already knew that the Witten genus vanished when $n$ was not divisible by $4$, since there are no modular forms of weight not divisible by $2$.}

\subsubsection{The mod-2 index}

That the indexes of time-reversal SQM models vanish in degrees $2$ and $6$ mod $8$ and are even in degree $4$ mod $8$ is compensated by a more refined ``mod-2 index,'' which is nontrivial in degrees $n=1$ and $2 \pmod 8$. We will review its construction because it already measures some torsion in the space of SQFTs and because a variation of the mod-2 index appears when trying to understand indexes of noncompact SQM models.
We start with the case $n=1$. The even subalgebra of $\Cliff(1,\bR)$ is isomorphic to $\bR$, and $\Cliff(1,\bR)$ has a unique irreducible module, namely itself. We will call it $\bR^{1|1}$. The supersymmetric ground states $V$ of a degree-$1$ SQM model  is then isomorphic to $\bR^{a|a}$ for some nonnegative integer $a$. By definition, the \define{mod-2 index} of the SQM model is $a \pmod 2$.

Although the integer $a$ is not a deformation invariant of the degree-$1$ SQM model, the mod-2 index $a\pmod 2$ is. To explain why, we can repeat the logic from \S\ref{sec.noncompact} to reinterpret a deformation of a degree-$n$ SQM model as a mildly-noncompact degree-$(n+1)$ SQM model: promote the deformation parameter to a dynamical supersymmetric multiplet; the fermion in this multiplet contributes $+1$ to the degree of the model. The upshot is that any deformation of a degree-$n$ SQM model will add or subtract to the ground states $V$ some $\Cliff(n+1,\bR)$-module (thought of as a $\Cliff(n,\bR)$-module).

But the even subalgebra of $\Cliff(2,\bR)$ is isomorphic to $\bC$ thought of as a real algebra.%
\footnote{As a real superalgebra, $\Cliff(2,\bR)$ is isomorphic to a ``semidirect tensor product'' $\bC \rtimes \Cliff(1,\bR)$, where the odd generator of $\Cliff(1,\bR)$ acts by complex conjugation on $\bC$.}
Because $\bC$ is a field, $\Cliff(2,\bR)$ is irreducible as a module over itself. We will call this irreducible module $\bC^{1|1}$. It has even graded dimension when restricted to $\Cliff(1,\bR)$, since $\dim_\bR \bC$ is even, and so adding or subtracting it to $V = \bR^{a|a}$ will not change the value of $a \pmod 2$. This is why the mod-2 index of a degree-$1$ SQM model is a deformation invariant.

The same logic also defines a deformation-invariant mod-2 index of a degree-$2$ SQM model. 
The 
 ground states $V$ are isomorphic, as a $\Cliff(2,\bR)$-module, to $\bC^{a|a}$ for some integer $a$, and the \define{mod-2 index} is  $a \pmod 2$. A deformation will involve adding or subtracting from $V$ the underlying $\Cliff(2,\bR)$-module of some $\Cliff(3,\bR)$-module. The even subalgebra of $\Cliff(3,\bR)$ is isomorphic to the quaternion algebra $\bH$. Because $\bH$ is a skew field, $\Cliff(3,\bR)$ is irreducible as a module over itself. Call this irreducible module $\bH^{1|1}$. Since $\dim_\bC \bH$ is even, if you add or subtract some multiple of $\bH^{1|1}$ to $\bC^{a|a}$, you do not change the value of $a \pmod 2$.

The story repeats when $n = 1$ or $2 \pmod 8$, because the category of $\Cliff(n,\bR)$-supermodules depends only on the value of $n \pmod 8$.
What about when $n = 3$? The ground states $V$ for an SQM model are then isomorphic to $\bH^{a|a}$ for some nonnegative integer $a$, and so we may still contemplate a \define{mod-2 index} defined to be the value of $a \pmod 2$. But now this mod-2 index is not a deformation invariant because $\Cliff(4,\bR)$ is not irreducible over itself. In fact, both irreps of $\Cliff(4,\bR)$ restrict over $\Cliff(3,\bR)$ to copies of $\bH^{1|1}$, and so only the dataless ``$a \pmod 1$'' is a deformation invariant. The same is true when $n = 5$, $6$, or $7 \pmod 8$.

\subsubsection{Noncompact SQM models}

We turn now to the index of ``mildly noncompact'' SQM models. The definition of ``mild noncompactness'' mirrors \S\ref{sec.noncompact}: the SQM model $\cX$ should come with a local operator $\Phi$ parameterizing the noncompact direction; writing $\cY_p$ for the theory produced from $\cX$ by adding a fermion $\lambda$ and a fermionic superpotential $\lambda(\Phi-p)$, we demand that $\cY_p$ stabilizes in the limits $p \to \pm \infty$ to compact SQM models $\cY_\pm$.
For definiteness, we will first describe the case when $\cX$ has degree $n=4$, in which case each $\cY_p$ is an SQM model of degree $3$.

We lose no generality by assuming that as $p$ varies, the Hilbert space $\cH$ of $\cY_p$ is independent of $p$, with the only variation being in the choice of supersymmetry.%
\footnote{Indeed, one may always achieve this by adding massive modes.}
Since $\cY_p$ has degree $3$, $\cH$ is naturally a module for $\Cliff(3,\bR)$. Choose%
\footnote{There are two choices for this isomorphism, yet another manifestation of the sign ambiguity in the definition of the Witten genus.}
 an isomorphism $\Cliff(3,\bR) \cong \bH \otimes \Cliff(-1,\bR)$, and let $\gamma$ denote the generator of $\Cliff(-1,\bR)$.
The supersymmetry generator in $\cY_p$ is $G(p) = g(p) \gamma$, where $g(p)$ is a quaternionic matrix; the time-reversal structure for SQM models of degree $3$ requires that $g(p)$ be ``quaternionically self-adjoint,'' meaning that its eigenvalues live in $\bR \subset \bH$. Thus, after a $p$-dependent change of basis, the only thing varying with $p$ is the spectrum of $g(p)$, which by compactness is a discrete subset (with finite multiplicities) of $\bR$.

The ``index'' of a noncompact SQM model $\cX$ can be defined as the Ramond partition function $Z_R(\cX) = \mathrm{Tr}_\cH (-1)^F \cdots$, but the noncompactness means that this partition function will depend nontrivially on the length of the worldline torus. The limit $\bar\tau \to -i\infty$ corresponds to the IR limit of $Z_R(\cX)$, which merely counts supersymmetric ground states. We will use the term ``index'' to mean this IR limit.

If the limits $g(\pm\infty) = \lim_{p \to \pm \infty} g(p)$ have no kernel, then the index (in the IR sense) of $\cX$ is relatively easy to compute: it counts with signs the number of eigenvalues of $g$ that cross~$0$, times a factor of 2 coming from the quaternionic nature of degree-$3$ and degree-$4$ SQM models. Indeed, to say that the limits $g(\pm\infty)$ have no kernel is to say that supersymmetry is spontaneously broken in these limits, and $\cX$ wasn't really ``noncompact'' at all, because it flows to a compact theory, and the factor of $2$ is the one coming from~(\ref{eqn.mofn}) when $n=4$.

If, on the other hand, supersymmetry is not spontaneously broken in the boundary theories $\cY_\pm$, then the index of $\cX$ receives a fractional contribution from each eigenvalue that lands on $0$ in the limits $p \to \pm \infty$. After multiplying by the factor of $2$ coming from~(\ref{eqn.mofn}), we find that the index of $\cX$ is odd depending on the number of supersymmetric ground states in the boundary theories $\cY_\pm$. And this number is exactly the non-deformation-invariant mod-2 index of $\cY_\pm$!

The same argument holds whenever $\cX$ has degree $n = 4 \pmod 8$. When $n = 0 \pmod 8$, a similar argument holds without a factor of $2$. To give a unified formula, we complexify and strip off the spectator fermions. Then, after complexifying, the supersymmetric ground states in $\cY_\pm$ form a vector space isomorphic to $\bC^{a|a}$, with $a \in m\bZ$, and the mod-2 index is $\frac a m \pmod 2$. We will call this number $a$ the ``bosonic index'' of $\cY_\pm$.
The end result is:
\begin{equation}\label{eqn.mod2index.n}
\operatorname{Index}(\cX) \in \frac 1 2 (\text{bosonic index of }\cY_+) - \frac 1 2 (\text{bosonic index of }\cY_-) + m\bZ ,
\end{equation}
where the degree $n$ of $\cX$ is divisible by $4$ and $m$ depends on $n$ via (\ref{eqn.mofn}).

In the non-IR, the more general statement identifies the failure of $Z_R(\cX)$ to be integral with a version of the $\eta$ invariant of \cite{MR0331443}.

\subsection{The invariant}\label{sec.invariant}

Summarizing the previous two sections, let $\cB$ be a compact SCFT%
\footnote{If one cares about an SQFT $\cB$ that is not initially conformal, then it should be RG-flowed to an SCFT before proceeding.
Without this step, one would also need anomaly equations encoding the dependence of $Z_{RR}(\cF)$ on the size of the worldsheet torus. As with the holomorphic anomaly, there is no ``bulk'' contribution to this dependence, but there can be a contribution from the ``boundary'' $\cB$. Indeed, if $\cB$ were not conformal, then the one-point function $g$ would include a size dependence. These size dependences are fully controlled by the Zamolodchikov c-theorem. 
One can confidently guess the IR limit of UV SQFTs with enough symmetry, and all examples in this paper are already SCFTs.
For instance, the examples in Section~\ref{sec.S3} are the expected IR limits of the round $S^3$ with various B-fields.}
 with gravitational anomaly $2(c_R - c_L) = n-1$, where $n$ is divisible by $4$. Suppose that $\cB$ is nullhomotopic. Then we can use the nullhomotopy to build an SQFT $\cF$ with gravitational anomaly $n$ with one noncompact direction and boundary $\cB$. Let $\hat{f}(\tau,\bar\tau)$ denote the Witten genus of $\cF$. Then:
\begin{enumerate}
  \item $\hat{f}$ is real-analytic modular of weight $(\frac n 2,0)$. It solves a \define{holomorphic anomaly equation}
  \begin{equation}\label{eqn.f1}
    \sqrt{-8\tau_2} \frac\partial{\partial \bar\tau} \hat f = g(\tau,\bar\tau)
  \end{equation}
  where
  \begin{equation}\label{eqn.g1}
    g(\tau,\bar\tau) = (\text{torus one-point function of } (-1)^{\frac{n}{4}}{:}\psi_1\cdots\psi_{n-1}\bar{G}{:}  \mbox{ in $\Fer(n-1) \otimes \cB$}).
  \end{equation}
  \item The holomorphic part $f(\tau) = \lim_{\bar\tau \to -i\infty} \hat{f}(\tau,\bar\tau)$ has $q$-expansion
  \begin{equation}\label{eqn.f2}
     f \in f_2(q) + m\bZ(\!(q)\!)
  \end{equation}
  where $m$ depends on $n$ via (\ref{eqn.mofn}) and
  \begin{equation}\label{eqn.g2}
    f_2(q) = \frac12 \bigl(\text{bosonic index of } \cB[S^1]\bigr).
  \end{equation}
\end{enumerate}
In particular, if $\cB$ is nullhomotopic, then there is a function $\hat f$ solving equations (\ref{eqn.f1}) and (\ref{eqn.f2}).

Conversely, 
suppose we suspect that $\cB$ is not nullhomotopic.
Equations (\ref{eqn.g1}) and (\ref{eqn.g2}) depend on $\cB$ alone. 
We can encode the data of $g$ by finding some (real-analytic modular of weight $(\frac n 2,0)$) solution $\hat{f}_1$ to (\ref{eqn.f1}). There is an ambiguity in the choice of solution, given precisely by the holomorphic modular forms of weight $\frac n 2$. Let $f_1$ denote the holomorphic part of $\hat{f}_1$. Then $g$, and hence $\cB$, determines the class of $f_1$ in
$$ [f_1] \in \frac{\bC(\!(q)\!)}{\MF_{n/2}}.$$
The theory $\cB$ determines the $q$-series $f_2 \in \frac m 2 \bZ(\!(q)\!)$ exactly, but we will use only its class in
$$ [f_2] \in \frac{\bC(\!(q)\!)}{m\bZ(\!(q)\!)}.$$

As we have seen already, neither of these classes is separately a deformation invariant of $\cB$. But we claim that the class
$$ [f_1] - [f_2] \in \frac{\bC(\!(q)\!)}{ \MF_{n/2} + m\bZ(\!(q)\!)} = A_n $$
is a deformation invariant.\footnote{The class $[f_2]$ is 2-torsion by construction, and so the sign in $[f_1] - [f_2]$ is irrelevant.} Indeed, suppose that $\cB$ can be deformed to some other SQFT $\cB'$. Then we can build from the deformation a mildly noncompact SQFT $\cF$ with boundaries $\cB_- = \cB$ and $\cB_+ = \cB'$. Let $f_1$ and $f_2$ denote the $q$-series for $\cB$ as defined above, and $f_1'$ and $f_2'$ the corresponding $q$-series for $\cB'$. Then $f_1'$ and $f_1 + Z_{RR}(\cF)$ solve the same holomorphic anomaly equation, and so differ by an element of $\MF_{n/2}$, whereas $f_2'$ and $f_2 + Z_{RR}(\cF)$ differ by an element in $m\bZ(\!(q)\!)$. This verifies that 
$$ f_1 - f_2 = f_1' - f_2' \mod \MF_{n/2} + m\bZ(\!(q)\!).$$

Thus, to solve a puzzle like Puzzle~\ref{mainpuzzle} or the other related puzzles from \cite{WittenTMF}, it suffices to calculate this invariant $[f_1] - [f_2]$ for the possibly-nullhomotopic SQFT $\cB$. In Sections~\ref{sec.cigar} and \ref{sec.S3} we will study examples with $\cB$ of degree $3$ (so in the above notation $n=4$ and $m=2$) where
$$ f_1(q) - f_2(q) = \alpha q^0 \mod \MF_{2} + 2\bZ(\!(q)\!),$$
for some rational number $\alpha \in \bQ$.
Such examples $\cB$ are definitely not nullhomotopic if $\alpha \not\in 2\bZ$. Indeed, it suffices to show that $\alpha q^0 \not\in \MF_{2} + 2\bZ(\!(q)\!)$ for $\alpha \not\in 2\bZ$, or equivalently that no weight-2 modular form has $q$-expansion in $\alpha + 2\bZ(\!(q)\!)$. But a weight-2 modular form is determined by its polar part, which would have to be even, but then the constant term would also have to be even.

\subsection{Relation to secondary invariants of Bunke--Naumann} \label{sec.BN}

The $A_n$-valued invariant of SQFTs that we have constructed  is intentionally modelled on the ``secondary invariant of the Witten genus'' constructed in \cite{MR3278648}. That paper builds, for any $(n-1)$-dimensional manifold $M$ equipped with String structure%
\footnote{A String structure consists of a spin structure together with a trivialization  of the characteristic class on spin manifolds called $\frac{p_1}2$. Target-space String structures are required in order to cancel the anomalies when building $\cN=(0,1)$ sigma models \cite{MR796163}.}
 with~$n$ divisible by~$4$, an invariant valued in the group $A_n$ above (called $T_{n/2}$ in \cite{MR3278648}); they give topological, geometric, and analytic descriptions of the invariant. Their invariant is a cobordism invariant, meaning that string-cobordant manifolds receive the same value. It is in a precise sense a ``derived invariant'' of the topologically-defined Witten genus.
 
This secondary invariant has a simple purely homotopy-theoretic description (our exposition follows \cite{DBE2015}). Let $\mathrm{HMF}_\bullet$ denote ordinary (de Rham) cohomology with coefficients in the graded ring $\MF$ of complex-valued modular forms, graded so that $\MF_{n/2}$ has cohomological degree $-n$.%
\footnote{The minus sign comes from the using cohomological rather than homological degree.}
 Write $\mathrm{H}\bC(\!(q)\!)_\bullet$ for $4$-periodicized ordinary cohomology with $\bC(\!(q)\!)$-coefficients. Let $\mathrm{KO}_\bullet$ denote orthogonal K-theory, and write $\mathrm{KO}(\!(q)\!)_\bullet$ for $S^1$-equivariant KO. Finally, let $\mathrm{MString}_\bullet$ denote the Thom spectrum of String cobordism.%
 \footnote{We will not distinguish cohomology theories from their corresponding spectra --- the cohomology theory associated to $\mathrm{MString}_\bullet$ is usually called $\Omega^{\mathrm{String}}_\bullet$.}
  Then the Witten genus fits into a commutative square of spectra:
$$ \begin{tikzpicture}[anchor=base]
  \path (0,0) node (MString) {$\mathrm{MString}_\bullet$} 
       +(3,0) node (KO) {$\mathrm{KO}(\!(q)\!)_\bullet$}
       +(0,-2) node (HMF) {$\mathrm{HMF}_\bullet$}
       +(3,-2) node (HCq) {$\mathrm{H}\bC(\!(q)\!)_\bullet$};
  \draw[->] (MString) -- (KO);
  \draw[->] (MString) -- (HMF);
  \draw[->] (KO) -- (HCq);
  \draw[->] (HMF) -- (HCq);
\end{tikzpicture}$$
This square is not a homotopy pullback square. Instead, construct the homotopy fiber product $\mathrm{HMF}_\bullet \times^h_{\mathrm{H}\bC(\!(q)\!)_\bullet} \mathrm{KO}(\!(q)\!)_\bullet$, christened ``$\mathrm{KMF}_\bullet$'' in \cite{DBE2015}. Then we automatically find a map:
\begin{equation}\label{KMF-diagram}
 \begin{tikzpicture}[anchor=base, baseline=(KMF.base)]
  \path (0,0) node (MString) {$\mathrm{MString}_\bullet$} 
       ++(1.5,-1) node (KMF) {$\mathrm{KMF}_\bullet$}
       +(.5,-.5) node {$\ulcorner$}
       +(3,0) node (KO) {$\mathrm{KO}(\!(q)\!)_\bullet$}
       +(0,-2) node (HMF) {$\mathrm{HMF}_\bullet$}
       +(3,-2) node (HCq) {$\mathrm{H}\bC(\!(q)\!)_\bullet$};
  \draw[->] (MString) -- (KO);
  \draw[->] (MString) -- (HMF);
  \draw[->] (KMF) -- (KO);
  \draw[->] (KMF) -- (HMF);
  \draw[->] (KO) -- (HCq);
  \draw[->] (HMF) -- (HCq);
  \draw[->,dashed] (MString) -- (KMF);
\end{tikzpicture}
\end{equation}
A short calculation verifies that, for $n$ divisible by $4$,
$$ \pi_{n-1}\mathrm{KMF}_\bullet \cong A_n,$$
and so some (possible trivial)
 $A_n$-valued cobordism invariant of $(n-1)$-dimensional String manifolds is automatic. The challenge, solved in \cite{MR3278648}, is to describe the invariant in a useful way.

As emphasized in \cite{DBE2015}, following \cite{MR885560,MR970288}, the maps $\mathrm{MString}_\bullet \to \mathrm{HMF}_\bullet$ and $\mathrm{MString}_\bullet \to \mathrm{KO}(\!(q)\!)_\bullet$ have natural quantum field theoretic descriptions. Indeed, given a (compact Riemannian) $n$-dimensional String manifold $M$, there is an $\cN=(0,1)$ supersymmetric sigma model with target $M$, with gravitational anomaly $n$.%
\footnote{The String structure is used to cancel an anomaly in the construction of the sigma model \cite{MR796163}.}
  String cobordisms produce deformations of SQFTs \cite[\S3.4]{WittenTMF}. Such a sigma model has a partition function in $\MF$; this defines the map $\mathrm{MString}_\bullet \to \mathrm{HMF}_\bullet$. On the other hand, an SQFT compactified on $S^1$ determines a time-reversal $S^1$-equivariant minimally supersymmetric quantum mechanics model, and so a point in $\mathrm{KO}(\!(q)\!)_\bullet$.%
  \footnote{The KO-valued deformation-invariant of SQM models is used regularly in physics --- for instance, we used it in \S\ref{sec.int}. The main theorem in \cite{MR2648897} says that it is a complete invariant: two SQM models are in the same deformation class if and only if their KO-valued invariants agree, and this remains true even in families, so that the $E_\infty$ ring spectrum $\mathrm{SQM}_\bullet$ is homotopy-equivalent to~$\mathrm{KO}_\bullet$.}

Let $\mathrm{SQFT}_n$ denote the moduli space of SQFTs with gravitational anomaly $-n$. This is not (as of this writing) a mathematically well-defined moduli space, not least because there is not a sufficient mathematical definition of $(1+1)$-dimensional quantum field theory (and even if $\mathrm{SQFT}_n$ could be defined as a set, correctly topologizing it would be hard). That said, if $\mathrm{SQFT}_n$ were a well-defined moduli space, then the spaces $\mathrm{SQFT}_\bullet$ would fit together into an $E_\infty$ ring spectrum. The adjoint to the suspension map $\Sigma \mathrm{SQFT}_n \to \mathrm{SQFT}_{n+1}$ making $\mathrm{SQFT}_\bullet$ into a spectrum is the map $\mathrm{SQFT}_n \to \Omega\mathrm{SQFT}_{n+1}$ which turns an SQFT $\cF$ into the family of SQFTs $t \mapsto {\Fer(1)(t)} \otimes \cF$, parameterized by $t \in \bR \cup \{\infty\} \cong S^1$, where $\Fer(1)(t)$
consists of a single chiral fermion $\psi$ with supersymmetry $\psi \mapsto t$.%
\footnote{As a QFT, $\Fer(1)(t) = \Fer(1)$ is conformal, but the supersymmetry breaks conformal invariance.}
 This map $\mathrm{SQFT}_n \to \Omega\mathrm{SQFT}_{n+1}$ is homotopy invertible: the inverse compiles a family of SQFTs $p \mapsto \cB_p$ into an a-priori-noncompact SQFT~$\cF$ just as in \S\ref{sec.noncompact}, but $\cF$ is in fact compact since the elements of $\Omega\mathrm{SQFT}_{n+1}$ are the families such that both limits $\cB_{\pm}$ are null.

In summary, conditional on giving a mathematical definition of the spaces $\mathrm{SQFT}_n$, the sigma-model and compactification maps allow us to expand (\ref{KMF-diagram}) into:
\begin{equation}\label{SQFT-diagram}
 \begin{tikzpicture}[anchor=base,baseline=(KMF.base)]
  \path (0,0) node (MString) {$\mathrm{MString}_\bullet$} 
       ++(1.5,-1) node (SQFT) {$\mathrm{SQFT}_\bullet$}
       ++(1.5,-1) node (KMF) {$\mathrm{KMF}_\bullet$}
       +(.5,-.5) node {$\ulcorner$}
       +(3,0) node (KO) {$\mathrm{KO}(\!(q)\!)_\bullet$}
       +(0,-2) node (HMF) {$\mathrm{HMF}_\bullet$}
       +(3,-2) node (HCq) {$\mathrm{H}\bC(\!(q)\!)_\bullet$};
  \draw[->] (SQFT) -- (KO);
  \draw[->] (SQFT) -- (HMF);
  \draw[->] (KMF) -- (KO);
  \draw[->] (KMF) -- (HMF);
  \draw[->] (KO) -- (HCq);
  \draw[->] (HMF) -- (HCq);
  \draw[->,dashed] (SQFT) -- (KMF);
  \draw[->] (MString) -- (SQFT);
\end{tikzpicture}.\end{equation}
This produces in particular a $\pi_{n-1}\mathrm{KMF}_\bullet$-valued invariant of SQFTs with gravitational anomaly $n-1$.

In topology, one can construct a diagram identical to (\ref{SQFT-diagram}) except with the mathematically ill-defined spectrum $\mathrm{SQFT}_\bullet$ replaced by the mathematically (although non-geometrically) well-defined spectrum $\mathrm{TMF}_\bullet$ of ``topological modular forms.'' (For details on TMF, see for example \cite{MR3223024}.) Bunke and Naumann prove in \cite{MR3278648} that their geometrically-defined invariant of String manifolds does factor through $\mathrm{TMF}_\bullet$. This is no accident: building on \cite{MR992209}, Stolz and Teichner proposed in \cite{MR2079378,MR2742432} that indeed $\mathrm{SQFT}_\bullet$ and $\mathrm{TMF}_\bullet$ are homotopy equivalent, in which case the former provides a much-needed geometric model of the latter.

\section{Example: $S^1$ and the cigar} \label{sec.cigar}

\subsection{SQM with target $S^1$} \label{subsec.SQM}

The $\cN=(0,1)$ sigma model requires a String structure on the target manifold \cite{MR796163}: a \define{String structure} consists of a spin structure together with a trivialization of the characteristic class on spin manifolds called $\frac{p_1}2$; the choice of trivialization is interpreted in the quantum field theory as a B-field \cite{MR1748791}. To illustrate the dependence on the spin structure, we warm up by discussing $\cN=1$ supersymmetric quantum mechanics with target $S^1$.\footnote{$\cN=1$ SQM models require the target to be spin but not String.}

The classical degrees of freedom for $\cN=1$ mechanics with target $X$ consist of a boson~$x$ valued in~$M$ (solving a second-order equation of motion) and a single real fermion~$\xi$ valued in the tangent bundle $x^* \mathrm{T}_X$ (solving a first-order equation of motion). The classical phase space is the symplectic supermanifold $\mathrm{T}^*  X  \times_X \Pi \mathrm {T} X$, where $\Pi \mathrm {T}$ means the parity-reversed tangent bundle. The algebra of quantum observables is the canonical quantization of this space. The cotangent bundle $\mathrm{T}^*  X$ quantizes to the algebra of differential operators on $X$, and the odd tangent bundle $\Pi \mathrm {T} X$ quantizes to (the global sections of) a bundle of Clifford algebras $\Cliff(\mathrm{T}_X)$. This algebra does not depend on the spin structure. Rather, the spin structure appears when trying to decide the Hilbert space. The (bosonic) algebra of differential operators has a canonical Hilbert space: the $L^2$ functions on $X$. But to choose a Hilbert space for the Clifford algebra factor, we must use the spin structure to choose a bundle of $\Cliff(\mathrm{T}_X)$-modules.

In the $S^1$ case, the tangent bundle $\mathrm{T}_{S^1}$ is trivial, and so the quantum algebra of observables is $\operatorname{DiffOp}(S^1) \otimes \Cliff(1,\bR)$. Write $\bR^{1|1}$ for the irreducible $\Cliff(1,\bR)$-module.
The nonbounding, aka periodic, spin structure corresponds to the Hilbert space $\cH_{\text{nonbounding}} = L^2(S^1) \otimes \bR^{1|1}$.
The supersymmetry is diagonalized in the momentum representation
$$ \cH_{\text{nonbounding}} \cong L^2(\bZ)  \otimes \bR^{1|1} \cong \bigoplus_{p\in\bZ} \bR^{1|1}.$$
Write $\gamma \in \Cliff(1)$ for the odd generator and $\hat{p} = \frac\d{\d x}$ for the momentum operator. In the momentum representation, the supersymmetry acts on the $p$th direct summand by $p\gamma$.
The supersymmetric ground states are therefore the $p=0$ summand $\bR^{1|1}$. Thus the mod-2 index is $1 \in \bZ/2\bZ$. This being nonzero confirms that the nonbounding spin structure is in fact nonbounding.

For the bounding, aka antiperiodic, spin structure, the sections of the Hilbert space have half-integral Fourier modes, and so in the momentum representation
$$ \cH_{\text{bounding}} \cong \bigoplus_{p \in \bZ+\frac12} \bR^{1|1},$$
again with supersymmetry $p\gamma$. Since $p$ now ranges over $\bZ + \frac12$, there are no supersymmetric ground states --- supersymmetry is spontaneously broken --- and so the mod-2 index vanishes, as it must for a bounding spin structure.

\subsection{The $S^1$ sigma models}\label{subsec.S1}

We turn now to the $\cN=(0,1)$ sigma model with target $S^1$, with its two possible spin structures. Each spin structure on $S^1$ extends to a unique String structure.

\subsubsection{Nonbounding spin structure, aka the standard circle SCFT} \label{subsubsec.S1-nonbounding}

Recall that the bosonic sigma model unpacks to a quantum mechanics model in the loop space of $S^1$, which is
$$ \Loops(S^1) \cong S^1 \times \bZ \times \bR^\infty.$$
The $S^1$ factor records the basepoint of the loop, the $\bZ$ factor records the winding number, and the $\bR^\infty$ factor records the vibrations of the loop. In this quantum mechanics interpretation, the sigma model action unpacks to a quantum mechanics action with a quadratic potential energy. This is diagonalized by using the momentum representation for the $S^1$ factor and the position representation for the $\bZ$ factor; the $\bR^\infty$ factor becomes a stack of harmonic oscillators, and is not particularly interesting for the present discussion. The result is that
$$\cH_{\text{bosonic}} \cong L^2(\bZ^2) \otimes (\text{infinitely many harmonic oscillators}).$$

If we choose the nonbounding spin structure, the $\cN=(0,1)$ sigma model factors as product of an antichiral free fermion and  a standard bosonic sigma model with circle target, just as in the SQM case of \S\ref{subsec.SQM}:
$$\cH^{(0,1)}_{\text{nonbounding}} \cong \cH_{\text{bosonic}} \otimes \bR^{1|1} \otimes (\cdots) \cong \bigoplus_{n,w} \bR^{1|1} \otimes (\dots).$$

The states of the bosonic sigma model  carry left- and right-moving momenta 
\begin{equation}
a_0 = \frac{n}{R} + \frac{w R}{2}, \qquad \qquad \bar a_0 = \frac{w R}{2} -\frac{n}{R},
\end{equation}
for the free-boson zero-modes, where $R$ is the radius of the target circle and the momentum $n$ and winding $w$ quantum numbers are integral. 
The momenta fill in an even self-dual lattice of signature $(1,1)$.  

The zero-mode contributions to the energy and momenta of the states give $L_0 = \frac12 \left(\frac{n}{R} + \frac{w R}{2} \right)^2$ and $\bar L_0 = \frac12 \left(\frac{n}{R} - \frac{w R}{2} \right)^2$. 
Also, $\bar G_0$ is proportional to $\bar a_0$. For generic values of the radius $R$ there is a single copy of $\bR^{1|1}$ with $n=w=0$. For rational values of 
$R^2$ one may have more general ground states with $2n = R^2 w$ and $L_0 = \frac{2 n^2}{R^2}$. That gives a single copy of $\bR^{1|1}$ when $n=0$; two copies of $\bR^{1|1}$ when $n^2 >0$.
Thus the mod-2 index is radius-independent and equals $1 \in \bZ_2(\!(q)\!)$. This proves that the $S^1$ sigma model with nonbounding spin structure
is not nullhomotopic in the moduli space of SQFTs.

For later reference, we can add a spectator fermion $\psi$ and compute the torus one-point function of $\sqrt{-1} {:}\psi \bar{G}{:}$. In order to get a non-zero answer, we 
can turn on non-trivial fugacities $y_1$, $y_2$ for the momentum and winding symmetries of the theory. 

In order to fix our notations, we write the anti-chiral supercurrent as
\begin{equation*}
\bar G = \sqrt{-1} \bar \psi \bar \partial \phi
\end{equation*}
and anti-chiral stress-tensor 
\begin{equation*}
\bar T = -\frac12 (\bar \partial \phi)^2 - \frac12 \bar \psi \bar \partial \bar \psi.
\end{equation*}
The basic free field OPE for the free boson currents  
\begin{equation*}
\bar \partial \phi(\bar z) \bar \partial \phi(\bar w) \sim - \frac{1}{(\bar z - \bar w)^2}
\end{equation*}
and the anti-chiral free fermion superpartner
\begin{equation*}
\bar \psi(\bar z) \bar \psi(\bar w) \sim \frac{1}{\bar z - \bar w}
\end{equation*}
give the expected superconformal OPE:
\begin{equation*}
\bar G(\bar z) \bar G(\bar w)= \frac{1}{(\bar z - \bar w)^3} + \frac{2 T(\bar w)}{\bar z - \bar w}.
\end{equation*}

We thus have 
\begin{equation*}
\sqrt{-1} \psi \bar G = - \psi \bar \psi \bar \partial \phi
\end{equation*}
It is easy to see that the $\psi \bar \psi$ one-point function is 
\begin{equation}\label{eqn-ferbarfer}
\mathrm{Tr} \psi_0 \bar \psi_0 (-1)^F q^{L_0-\frac{1}{48}}\bar q^{L_0-\frac{1}{48}} =\sqrt{-1} |\eta(\tau)|^2 
\end{equation}
where the factor $|\eta(\tau)|^2$ comes from the trace over the Fock space of non-zero-modes, while the factor of $ \sqrt{-1}$ is the trace of the product of the zero-modes, i.e. 
the trace of $\gamma^1 \gamma^2$ in the irreducible module for $\Cliff(2,\bC)$. 

On the other hand, the $\sqrt{-1} \bar \partial \phi$ one-point function is 
\begin{equation*}
\mathrm{Tr}\, (-\bar a_0) (-1)^F q^{L_0-\frac{1}{24}}\bar q^{L_0-\frac{1}{24}} = \frac{1}{|\eta(\tau)|^2}\sum_{n,w \in \bZ}  \left(\frac{n}{R}- \frac{w R}{2} \right) y_1^n y_2^w q^{\frac{1}{2} (\frac{w R}{2} + \frac{n}{R})^2} \bar q^{\frac{1}{2} (\frac{w R}{2} - \frac{n}{R})^2}.
\end{equation*}
We conclude that the one-point function of $\sqrt{-1} {:}\psi \bar{G}{:}$ equals 
\begin{equation}\label{eqn-nonboundingS1}
\sum_{n,w \in \bZ} \left(\frac{w R}{2} - \frac{n}{R} \right) y_1^n y_2^w q^{\frac{1}{2} (\frac{w R}{2} + \frac{n}{R})^2} \bar q^{\frac{1}{2} (\frac{w R}{2} - \frac{n}{R})^2}.
\end{equation}

\subsubsection{Bounding spin structure, aka the exotic circle SCFT} \label{subsubsec.S1-bounding}

For the bounding spin structure, the fermion groundstate is taken to be anti-periodic along the target circle. In the sector with even winding number, the momentum lattice is shifted by $\frac12$ just as in the SQM case of \S\ref{subsec.SQM}. In sectors with odd winding number, the fermion ground-state must be anti-periodic as one goes around the space circle. This can be implemented by an extra shift by $\frac12$ of the momentum lattice. That means that the momentum 
lives in a shifted lattice $\bZ +\frac{w-1}{2}$. For generic values of the radius there is now no ground state at all, while for rational values of 
$R^2$ one may have more general ground states with $2n = R^2 w$ and $L_0 =  \frac{2 n^2}{R^2}$ which give two copies of $\bR^{1|1}$ for each possible value of $n^2$. Thus the mod-2 index is radius-independent and equals $0 \in \bZ_2(\!(q)\!)$. 

We can actually give a simple description of the full SCFT associated to the $S^1$ sigma model with bounding spin structure. 
We claim that it is the product of an anti-chiral free fermion and of an ``exotic'' free boson spin-CFT, based on an {\it odd}
self-dual lattice of signature $(1,1)$ (see also \cite{Karch:2019lnn} for a discussion of free boson spin-CFTs): 
\begin{equation}
a_0 = \frac{m}{2R} + \frac{w R}{2}, \qquad \qquad \bar a_0 = \frac{w R}{2} -\frac{m}{2R}
\end{equation}
The integers $(m,w)$ are either both even or both odd in the NS sector of the theory, while they have opposite parity in the 
R sector. The fermion number can be defined as $(-1)^F = (-1)^w$ or $(-1)^F = (-1)^m$ depending on the choice of overall fermionic parity 
of the Ramond sector. 

An alternative description of the theory is that of a twisted orbifold of a standard free boson CFT of radius $2R$ by a $\bZ_2$ translation symmetry, 
where the twist involves the non-trivial fermionic 2-cocycle for $\bZ_2$ \cite{GuWen2014}: in the language of \cite{GJFIII}, it is the fermionic theory 
whose bosonic neighbours are the standard free boson CFTs of radii $2R$ and $R$. 

We conclude that the one-point function of $\sqrt{-1} {:}\psi \bar{G}{:}$ equals 
\begin{equation}\label{eqn-boundingS1}
\sum_{m,w \in \bZ |m-w-1  \in 2\bZ} \left(\frac{w R}{2} - \frac{m}{2R} \right) (-1)^w y_1^{\frac{m}{2}} y_2^w q^{\frac{1}{2} (\frac{w R}{2} + \frac{m}{2R})^2} \bar q^{\frac{1}{2} (\frac{w R}{2} - \frac{m}{2R})^2}.
\end{equation}

\subsection{$T^3$ with Lie group framing}

Take $S^1$ with nonbounding String structure and multiply it by itself three times; the result is the 3-torus $T^3$ with String structure coming from the Lie group framing. As a spin manifold, this $T^3$ is bounding (it bounds a certain ``half K3 surface''), but it is nonbounding as a String manifold, and represents a class of exact order $2$ in TMF.

The (unflavoured) one-point function of $\bar G$ vanishes identically in the $T^3$ sigma model. So the class called $[f_1] \in \bC(\!(q)\!) / \MF_2$ in
\S\ref{sec.invariant} vanishes.
 But the class $[f_2]$ does not vanish. Rather, it is the cube of the mod-2 index  for the $S^1$ model, which we computed in \S\ref{subsec.S1} to be $1 \in \bZ_2(\!(q)\!)$. Thus, for the sigma model with target $T^3$ with Lie group framing,
$$ [f_2] \equiv 1 \in \frac{\bC(\!(q)\!) }{ 2\bZ(\!(q)\!)}.$$
It follows that the deformation-invariant class $[f] = [f_1] + [f_2]$ is
$$ [f] \equiv 1 \in A_4 = \frac{\bC(\!(q)\!) }{\MF_2 + 2\bZ(\!(q)\!)}, $$
which has exact order $2$.

In fact, $T^3$ with this String structure is string-cobordant to $S^3$ with B-field of strength $k=12$. We will calculate directly the invariant of $(S^3, k=12)$ in Section~\ref{sec.S3}, and see that it is equal to the invariant of $T^3$.

\subsection{The $\cN=(1,1)$ cigar sigma model}

The simplest test of our holomorphic anomaly equation~(\ref{eqn.holomorphicanomaly}) is to consider the elliptic genus of an $\cN=(0,1)$ sigma model whose target is the 
\define{cigar}. This is a K\"ahler manifold which coincides with $\bC$ as a complex manifold but is endowed with a K\"ahler metric which 
asymptotes to the flat metric on $\bR \times S^1$ with some radius $R$. Hence the ``boundary theory'' is an $\cN=(0,1)$ circle sigma model. 

The one-point function of $\bar{G}$ in the $(0,1)$ circle sigma model unfortunately vanishes: $\bar{G} = \bar \psi \bar \partial \phi$ is the product of the
anti-chiral fermion and the anti-chiral $U(1)$ current for translations of the circle, whose torus one-point function vanishes for symmetry reasons. 
There is a simple way to produce a nontrivial holomorphic anomaly equation: we can look at the flavoured elliptic genus, 
using the rotational symmetry of the cigar. 

The elliptic genus of the $\cN=(1,1)$ version of this theory is very well studied \cite{Eguchi:2006tu,MR2821103,Ashok:2011cy,murthy13}, and is already a useful example. It can be computed by localization techniques \cite{Ashok:2013pya}, as the theory is actually endowed with $(2,2)$ SUSY and can be obtained from RG flow of a $U(1)$ gauge theory \cite{Hori:2001ax}: the cigar geometry is a Kahler quotient of $\bC \times \bC^*$ by $U(1)$, acting as rotations of the first factor and translations of the second factor. 
 The elliptic genus is:
\begin{equation}\label{eqn.cigar1}
Z_{RR}( \mathrm{cig}^{(1,1)})[\xi;\tau] = g^{2} \int_{\bR^2} \d u_1 \d u_2 \frac{\theta(u_1 + \tau u_2 -\xi;\tau)}{\theta(u_1 + \tau u_2;\tau)} e^{- \frac{\pi g^2}{\tau_2} (u_1 + \tau u_2 + \frac{\xi}{g^2})(u_1 + \bar \tau u_2 + \frac{\xi}{g^2})}
\end{equation}
The $U(1)$ symmetry with fugacity $\xi$ is the chiral R-symmetry. It acts both on the chiral fermions and on the cigar sigma model, with a specific charge. 
We will often use the notation $u = u_1 + \tau u_2$. 

The cancellation of gauge and mixed anomalies is manifested in the invariance of the integrand under the modular transformation 
$\tau \to - \frac{1}{\tau}$, $u \to \frac{u}{\tau}$, $z \to \frac{z}{\tau}$, up to a factor of $\exp 2 i \pi (\frac12 + \frac{1}{g^2}) \frac{z^2}{\tau}$ encoding the 
flavour 't Hooft anomaly coefficient $1 + \frac{2}{g^2}$. 

\subsubsection{Integrality of the $q$-expansion}\label{subsec.nonboundingS1integrality}
As a check of our normalization, we should verify the integrality of the $q$ expansion of the holomorphic part of the elliptic genus. 
The $\bar \tau \to - i \infty$ limit of (\ref{eqn.cigar1}) gives 
\begin{equation}\label{eqn.cig11.d}
Z_{RR}^{hol}( \mathrm{cig}^{(1,1)})[\xi;\tau] = g^{2} \int_{\bR \times \bR} \d u_1 \d u_2 \frac{\theta(u_1 + \tau u_2-\xi;\tau)}{\theta(u_1 + \tau u_2;\tau)} e^{2 i \pi g^2 (u_1 + \tau u_2 + \frac{\xi}{g^2})u_2 }
\end{equation}
We can bring $u_1$ to the range $[0,1]$:
\begin{equation*}
Z_{RR}^{hol}( \mathrm{cig}^{(1,1)})[\xi;\tau] = \sum_{t \in \bZ}  g^{2} \int_{[0,1] \times \bR} \d u_1 \d u_2 \frac{\theta(u_1 + \tau u_2-\xi;\tau)}{\theta(u_1 + \tau u_2;\tau)}e^{2 i \pi g^2 (t+ u_1 +  \tau u_2 + \frac{\xi}{g^2})u_2 },
\end{equation*}
which happens to be the Poisson resummation of a simpler sum:
\begin{equation}\label{eqn.cig11.e}
Z_{RR}^{hol}( \mathrm{cig}^{(1,1)})[\xi;\tau] =\sum_{s \in \bZ} \int_0^1\d u_1 \frac{\theta(u_1 + \frac{\tau}{g^2} s-\xi;\tau)}{\theta(u_1 + \frac{\tau}{g^2} s;\tau)}e^{2 i \pi s (u_1  + \frac{\tau}{g^2} s + \frac{\xi}{g^2}) }
\end{equation}
The integral in (\ref{eqn.cig11.e}) computes the Fourier coefficient of $\frac{\theta(z-\xi;\tau)}{\theta(z;\tau)}$ along a circle 
positioned at $\frac{\tau}{g^2} s$. The Fourier coefficients of such a meromorphic Jacobi form are integral series in $x$ and $q$ which depend sensitively on the integral part of $\frac{s}{g^2}$. If we denote them as $f_s^{(\frac{s}{g^2})}(\xi;\tau)$, we get 
\begin{equation}
Z_{RR}^{hol}( \mathrm{cig}^{(1,1)})[\xi;\tau] =\sum_{s \in \bZ} f_s^{(\frac{s}{g^2})}(\xi;\tau) e^{2 i \pi s \frac{\xi}{g^2} }.
\end{equation}

We should remark that a lot of the formulae simplify when $g^2$ is rational. In particular, for integer $g^2 = k$ one obtains so-called Appell--Lerche sums. 

\subsubsection{The holomorphic anomaly equation}\label{sec.cig11anomaly}
The holomorphic anomaly was computed in the literature (see e.g. \cite{murthy13}). We will redo the calculation here to make sure we keep track carefully of the overall normalization. 
We can take the $\bar \tau$ derivative directly in the integral formula (\ref{eqn.cigar1}). 

It is easy to see that the exponential in the integrand in (\ref{eqn.cigar1}) is annihilated by $\frac{\partial}{\partial\bar{\tau}}  - \frac{i}{2 \pi g^2}\frac{\partial^2}{\partial\bar{u}^2}$, where $\frac{\partial}{\partial\bar{u}} = \frac{\tau \partial_{u_1} - \partial_{u_2}}{\tau - \bar \tau}$. As a result we have 
\begin{equation}\label{eqn.cig11.a}
\frac{\partial}{\partial\bar{\tau}} Z_{RR}( \mathrm{cig}^{(1,1)})[\xi;\tau] = \frac{i}{2\pi}  \int_{\bR^2} \d u_1 \d u_2 \frac{\theta(u-\xi;\tau)}{\theta(u;\tau)} \frac{\partial^2}{\partial\bar{u}^2} \left[e^{- \frac{\pi g^2}{\tau_2} (u + \frac{\xi}{g^2})(\bar u + \frac{\xi}{g^2})} \right]
\end{equation}
where $u = u_1 + \tau u_2$ and thus $\frac{\partial u}{\partial\bar{u}} =0$. 

Then we can integrate by parts to get 
\begin{equation*}
\frac{\partial}{\partial\bar{\tau}} Z_{RR}( \mathrm{cig}^{(1,1)})[\xi;\tau] = \frac{1}{2\pi i}  \int_{\bR^2} \d u_1 \d u_2 \frac{\partial}{\partial\bar{u}}  \left[\frac{\theta(u-\xi;\tau)}{\theta(u;\tau)}\right] \frac{\partial}{\partial\bar{u}}  \left[e^{- \frac{\pi g^2}{\tau_2} (u + \frac{\xi}{g^2})(\bar u + \frac{\xi}{g^2})}\right] .
\end{equation*}
As $\frac{\partial}{\partial\bar{u}} \frac{1}{u} = \frac{\pi}{\tau_2} \delta(u_1)\delta(u_2)$, 
the derivative picks the poles of $\frac{\theta(u-\xi;\tau)}{\theta(u;\tau)}$ at $u = n \tau + m$ and converts the integral into a sum 
\begin{equation*}
\frac{\partial}{\partial\bar{\tau}} Z_{RR}( \mathrm{cig}^{(1,1)})[\xi;\tau]  = \frac{g^2}{4 \tau^2_2} \frac{\theta(\xi;\tau)}{\eta(\tau)^3} \sum_{n,m \in \bZ}  x^{-n} (n \tau + m + \frac{\xi}{g^2}) e^{- \frac{\pi g^2}{\tau_2} (n \tau + m + \frac{\xi}{g^2})(n\bar \tau +m + \frac{\xi}{g^2})} ,
\end{equation*}
where we used 
\begin{equation*}
\theta(\xi + n \tau + m;\tau) = (-1)^{n+m} q^{-\frac{n^2}{2}} x^{-n}\theta(\xi;\tau).
\end{equation*}
Poisson resummation in $m$ finally gives 
\begin{equation}\label{eqn.cigar11.b}
\sqrt{-8 \tau_2} \frac{\partial}{\partial\bar{\tau}} Z_{RR}( \mathrm{cig}^{(1,1)})[\xi;\tau] =  \frac{\theta(\xi;\tau)}{\eta(\tau)^3} \sum_{n,s \in \bZ}  x^{\frac{s}{g^2} +n} \frac{1}{\sqrt{2}} (g n - \frac{s}{g}) q^{\frac{1}{4}(n g +s/g)^2} \bar q^{\frac{1}{4}(n g -s/g)^2}. \end{equation}

We recognize on the right-hand side of (\ref{eqn.cig11.a}) the sum over momenta and winding of an $S^1$ sigma model. If we define $R^2 = 2 g^2$, rename the summation variables and 
reshuffle some factors of $\eta$, we can write 
\begin{equation}\label{eqn.cigar11.c}
\sqrt{- 8 \tau_2} \frac{\partial}{\partial\bar{\tau}} \eta(\tau)^2 Z_{RR}( \mathrm{cig}^{(1,1)})[\xi;\tau] = \frac{\theta(\xi;\tau)}{\eta(\tau)} \sum_{n,w \in \bZ} (\frac{w R}{2} - \frac{n}{R}) x^{\frac{2 n}{R^2} +w}   q^{\frac{1}{2} (\frac{w R}{2} + \frac{n}{R})^2} \bar q^{\frac{1}{2} (\frac{w R}{2} - \frac{n}{R})^2} \end{equation}

The right-hand side of (\ref{eqn.cigar11.c}) is precisely the product of the flavoured torus partition function of $\Fer(2)$ and the flavoured torus one-point function (\ref{eqn-nonboundingS1}) of $\psi \bar G$ in the sigma model with nonbounding $S^1$ target and spectator fermion $\psi$. The left-hand side is the holomorphic anomaly of the $(1,1)$ cigar theory with two extra spectator fermions. It is not a contradiction to find the nonbounding $S^1$ here: the nonbounding $S^1$ can become a boundary when combined non-trivially with two extra chiral fermions.

\subsection{The $\cN=(0,1)$ cigar sigma model}

The elliptic genus for the $\cN=(0,1)$ sigma model (which actually has $\cN=(0,2)$ supersymmetry) can be computed in an analogous manner, though 
one needs to adjust a bit the formulae available in the literature. The main subtlety is the cancellation of gauge anomalies in the 
$\cN=(0,1)$ gauge theory, since, unlike in the $\cN=(1,1)$ case, this cancellation depends on a String structure on the target manifold.
For the $\cN=(0,1)$ cigar sigma model, our localization formula for the elliptic genus is:
\begin{equation}
Z_{RR}(\mathrm{cig})[\xi;\tau] = g^{2}  \int_{\bR^2} \d u_1 \d u_2  \frac{\eta(\tau)^3}{\theta(u+(1+g^{-2}) \xi;\tau)} e^{- \frac{\pi g^2}{\tau_2}  (u + \frac{\xi}{g^2})(\bar u + \frac{\xi}{g^2})- \frac{\pi}{2\tau_2}(u + \frac{\xi}{g^2})^2}
\end{equation}
The factor $e^{- \frac{\pi}{2\tau_2}(u + \frac{\xi}{g^2})^2 }$ is a correction which seems to be missing from the literature.
We take $\xi$ to be real. 

The overall normalization can be fixed by the observation that the limit $g^2 \to \infty$ should bring the theory back to the $\cN=(0,1)$ sigma model with target $\bR^2$. 
Indeed, in this limit the exponent goes to a $g^{-2} \delta(u_1) \delta(u_2)$ and the elliptic genus goes to $\frac{\eta(\tau)^3}{\theta(\xi;\tau)}$, as it should. 

\subsubsection{Integrality of the $q$-expansion}

The holomorphic part of the elliptic genus is 
\begin{equation}
Z^{hol}_{RR}(\mathrm{cig})[\xi;\tau] = g^{2} \int_{\bR \times \bR} \d u_1 \d u_2 \frac{\eta(\tau)^3}{\theta(u+(1+g^{-2}) \xi;\tau)}  e^{2 i \pi g^2  (u_1 + \tau u_2 + \frac{\xi}{g^2})u_2}
\end{equation}
which can be manipulated as in \S\ref{subsec.nonboundingS1integrality}:
\begin{multline}
Z^{hol}_{RR}(\mathrm{cig})[\xi;\tau] =  g^{2} \sum_{m \in \bZ} \int_{[0,1]\times \bR} \d u_1 \d u_2  \frac{\eta(\tau)^3}{\theta(u+(1+g^{-2}) \xi;\tau)}(-1)^m e^{2 i \pi g^2  (u_1 + m +\tau u_2 + \frac{\xi}{g^2})u_2}
\end{multline}
and then 
\begin{multline}
Z^{hol}_{RR}(\mathrm{cig})[\xi;\tau] =   \sum_{n \in \bZ} \int_{[0,1]} \d u_1 \frac{\eta(\tau)^3}{\theta(u_1+\tau (n+\frac12)\frac{1}{g^2} +(1+g^{-2}) \xi;\tau)}  \\ \times (-1)^m e^{2 i \pi  (u_1 +\tau (n+\frac12)\frac{1}{g^2} + \frac{\xi}{g^2})(n+\frac12)}
\end{multline}
which is doing Fourier transforms along a circle located at $\tau (n+\frac12)\frac{1}{g^2} +(1+g^{-2}) \xi$ of $ \frac{\eta(\tau)^3}{\theta(z;\tau)}$. Again, the expansion is manifestly integral. 

\subsubsection{The holomorphic anomaly equation} 
The calculation of the holomorphic anomaly can proceed as in \S\ref{sec.cig11anomaly}. The exponential in the integrand is annihilated by the usual heat operator
$\frac{\partial}{\partial\bar{\tau}}  - \frac{i(1+2 g^2)}{4 \pi g^4}\frac{\partial^2}{\partial\bar{u}^2}$, with $\frac{\partial}{\partial\bar{u}} = \frac{\tau \partial_{u_1} - \partial_{u_2}}{\tau - \bar \tau}$.
Then we can integrate by parts to get 
\begin{multline}
\frac{\partial}{\partial\bar{\tau}}  Z_{RR}( \mathrm{cig})[\xi;\tau]   =  \frac{1}{2\pi i} (1+\frac{1}{2 g^2})  \int_{\bR^2} \d u_1 \d u_2\frac{\partial}{\partial\bar{u}}  \left[\frac{\eta(\tau)^3}{\theta(u+(1+g^{-2}) \xi;\tau)}\right] \\ \times \frac{\partial}{\partial\bar{u}}   \left[ e^{- \frac{\pi g^2}{\tau_2}  (u + \frac{\xi}{g^2})(\bar u + \frac{\xi}{g^2})- \frac{\pi}{2\tau_2}(u + \frac{\xi}{g^2})^2}\right] 
\end{multline}
This picks the poles of $\frac{\eta(\tau)^3}{\theta(u+(1+g^{-2}) \xi;\tau)}$ at $u = n \tau + m-(1+g^{-2}) \xi$ and converts the integral into a sum 
\begin{multline}
 \frac{\partial}{\partial\bar{\tau}}Z_{RR}( \mathrm{cig})[\xi;\tau] =- (g^2+\frac{1}{2})  \frac{1}{4\tau_2^2} \sum_{n,m}(-1)^{n+m} q^{\frac{n^2}{2}} x^{- (1+g^{-2})n} (n \tau + m- \xi )
\\ \times \left[ e^{- \frac{\pi g^2}{\tau_2}  (n \tau + m- \xi )(n \bar \tau + m-\xi )- \frac{\pi}{2\tau_2}(n \tau + m- \xi )^2}\right] 
\end{multline}
where we used 
\begin{equation*}
\theta(\xi + n \tau + m;\tau) = (-1)^{n+m} q^{-\frac{n^2}{2}} x^{-n}\theta(\xi;\tau)
\end{equation*}
Poisson resummation in $m$ finally gives a nice expression. Defining $\gamma^2 = g^2 + \frac12$ we have  
\begin{multline*}
\sqrt{-4\tau_2} \frac{\partial}{\partial\bar{\tau}} \eta(\tau)^2 Z_{RR}( \mathrm{cig})[\xi;\tau] 
\\ = - \sum_{n,s \in \bZ} (-1)^n x^{\frac12 -s-n -\frac{1}{g^2} n} \frac{1}{2} (\gamma n - \frac{s+\frac{n}{2}-\frac12}{\gamma}) q^{\frac{1}{4}(\gamma n +\frac{s+\frac{n}{2}-\frac12}{\gamma})^2} \bar q^{\frac{1}{4}(\gamma n - \frac{s+\frac{n}{2}-\frac12}{\gamma})^2} ,
\end{multline*}
or, equivalently, with $R=\gamma$, 
\begin{multline}\label{eqn.cig12.b}
\sqrt{-8 \tau_2} \frac{\partial}{\partial\bar{\tau}} \eta(\tau)^2 Z_{RR}( \mathrm{cig})[\xi;\tau] 
\\ = \sum_{\substack{w,m \in \bZ, \\w+m =  \mathrm{odd}}} \left(\frac{m}{2R}-\frac{w R}{2} \right)  (-1)^w x^{-\frac{w+m}{2} -\frac{1}{g^2} w}q^{\frac{1}{2}(\frac{m}{2R}+\frac{w R}{2} )^2} \bar q^{\frac{1}{2}(\frac{m}{2R}-\frac{w R}{2} )^2} .
\end{multline}
The right-hand side of (\ref{eqn.cig12.b}) is nothing but the one-point function of $\psi \bar{G}$ of the circle theory with bounding spin structure (and a spectator fermion). The form of the momentum and winding lattice and the extra $(-1)^w$ factor are explained in \S\ref{subsec.S1}.

\section{Example: $S^3$ with WZW coupling $k$}\label{sec.S3}

The $\cN=(0,1)$ sigma model with target $S^3$ and WZW coupling $k$ is expected to 
 flow in the IR to an $\mathrm{SU}(2)$ WZW model, which is the same as a 
bosonic WZW model at level $\kappa = k-1$ together with three anti-chiral free fermions \cite{WittenTMF}. 
In the special case $k=1$, the bosonic WZW model at level $\kappa=0$ is the trivial theory, and the model flows to the free-fermion SCFT from Puzzle~\ref{mainpuzzle}. 
In all cases, the SQM produced by compactifying of the model on $S^1$ has spontaneous supersymmetry breaking \cite[\S4]{WittenTMF}, and so to compute our invariant we must only compute the $q$-series called $f_1$ in \S\ref{sec.invariant}.
To warm up, we address the $k=1$ case in \S\ref{subsec.S31}. The general case is in \S\ref{subsec.S3k}.

\subsection{A warm-up: $k=1$}\label{subsec.S31}
The anti-chiral stress tensor in the $\overline{\Fer}(3)$ theory is 
\begin{equation*}
\bar T = - \frac12 \bar \psi_a \bar \partial \bar \psi_a
\end{equation*}
The supercurrent can be taken to be 
\begin{equation}\label{eqn.barG1}
\bar G =\sqrt{-1} \bar \psi_1 \bar \psi_2 \bar \psi_3
\end{equation}
so that 
\begin{equation*}
\bar G(\bar z) \bar G(\bar w)= \frac{1}{(\bar z - \bar w)^3} + \frac{2 T(\bar w)}{\bar z - \bar w}.
\end{equation*}

Then we have that the one-point function of $-\psi_1 \psi_2 \psi_3 \bar G$ equals $|\eta(\tau)|^6$,
though the sign is somewhat conventional.

The function 
\begin{equation}\label{F1-formula}
F_1(\tau) = -\frac{1}{24} + \sum_{n=1}^{\infty} n\, \frac{q^n}{1-q^n}+\sum_{n=1}^{\infty} (-1)^{n-1} n \,\frac{q^{\frac{n(n+1)}{2}} }{1-q^n}
\end{equation}
has a modular, non-holomorphic completion $\hat F_1$ of weight 2 which satisfies 
\begin{equation}\label{F1-holomorphicanomaly}
\sqrt{- 8 \tau_2} \frac{\partial \hat F_1}{\partial \bar \tau} = \frac12 |\eta|^6.
\end{equation}
and thus $2 \hat F_1$ solves our holomorphic anomaly equation.

This shows immediately that the invariant for the $k=1$ case is $-\frac1{12} \pmod 2$, as expected! In particular, 
only 24 copies of $\overline{\Fer}(3)$ can be null-homotopic. 

In order to verify these assertions, first proven by 
the first named author
in collaboration with D. Zagier, 
one may employ an integral formula for $\hat F_1$:
\begin{equation}\label{F1-integralformula}
\hat F_1 = \frac{1}{8 \pi^2} \int_0^1 \int_0^1 \wp(u_1 + \tau u_2,\tau) H_1(u_1, u_2; \tau, \bar \tau)\, \d u_1 \d u_2 ,
\end{equation}
where $\wp(u_1 + \tau u_2,\tau)$ is the Weierstrass function and 
\begin{equation}\label{F1-formulaforH1}
H_1(u_1, u_2; \tau, \bar \tau) \equiv \sum_{n,m \in \bZ} e^{2 \pi i (n u_2 - m u_1)} (-1)^{n + m + nm} e^{- \frac{\pi}{2 \tau_2}|m \tau + n|^2},
\end{equation}
which can be Poisson resummed to 
\begin{equation*}
H_1(u_1, u_2; \tau, \bar \tau) \equiv  \sqrt{2\tau_2}|\theta(u_1 + \tau u_2,\tau)|^2 e^{- 2 \pi \tau_2 u_2^2} .
\end{equation*}

In the $\bar \tau \to - \infty$ limit, we recover 
\begin{equation*}
F_1 = \frac{1}{8 \pi^2} \int_0^1 \int_0^1 \wp(u_1 + \tau u_2,\tau) \sum_{n,m \in \bZ} e^{2 \pi i (n u_2 - m u_1)} (-1)^{n + m} e^{i \pi m^2 \tau }\d u_1 \d u_2 ,
\end{equation*}
which is the Poisson resummation of  
\begin{equation*}
F_1 = \frac{1}{8 \pi^2} \int_0^1 \wp(u_1 + \frac{\tau}{2} ,\tau) \sum_{m \in \bZ} e^{-2 \pi i m u_1} (-1)^{m} e^{i \pi m^2 \tau }\d u_1 .
\end{equation*}
We can use the Fourier expansion 
\begin{equation*}
\frac{1}{(2 \pi i)^2} \wp(\xi,\tau) = -2 G_2(\tau) + \sum_{n\in \bZ | n\neq 0} \frac{n}{1-q^n}x^n
\end{equation*}
along that circle to get the expected
\begin{equation}\label{F1-formula-redux}
F_1 = G_2 -  \sum_{m > 0 }  \frac{m}{1-q^m} (-1)^{m} e^{i \pi m(m+1) \tau }
\end{equation}
where 
\begin{equation*}
G_2 = -\frac{1}{24} + \sum_{n=1}^{\infty} n\, \frac{q^n}{1-q^n}
\end{equation*}
is the second Eisenstein series. Formula (\ref{F1-formula}) is a restatement of (\ref{F1-formula-redux}).

At finite $\bar \tau$ we can still use the Fourier expansion of  $\wp(\xi,\tau)$ to get 
\begin{align*}
\hat F_1 & = G_2(\tau)-\frac{1}{2} \int_0^1  \sum_{n,m \in \bZ| m\neq 0}  \frac{m}{1-q^m}e^{2 \pi i (m \tau +n) u_2 } (-1)^{n + m + nm} e^{- \frac{\pi}{2 \tau_2}|m \tau + n|^2} \d u_2 ,
\\
& = G_2(\tau)-\frac{i}{4 \pi}   \sum_{n,m \in \bZ| m \neq 0} (-1)^{n + m + nm} \frac{m}{m \tau + n}  e^{- \frac{\pi}{2 \tau_2}|m \tau + n|^2}.
\end{align*}
This satisfies a holomorphic anomaly equation
\begin{equation*}
\partial_{\bar \tau} \hat F_1 = \frac{1}{16 \tau_2^2}   \sum_{n,m \in \bZ} (-1)^{n + m + nm} m(m \tau + n)  e^{- \frac{\pi}{2 \tau_2}|m \tau + n|^2},
\end{equation*}
which is Poisson resummed to the desired (\ref{F1-holomorphicanomaly}).

We can obtain the same result by starting from the heat equation
\begin{equation*}
\partial_{\bar \tau} H_1(u_1, u_2; \tau, \bar \tau)= \frac{i}{4 \pi} \partial_{\bar u}^2 H_1(u_1, u_2; \tau, \bar \tau)
\end{equation*}
with $\frac{\partial}{\partial\bar{u}} = \frac{\tau \partial_{u_1} - \partial_{u_2}}{\tau - \bar \tau}$ and $H_1$ from (\ref{F1-formulaforH1}).
Integrating by parts in (\ref{F1-integralformula})
then gives 
\begin{equation*}
\partial_{\bar \tau} \hat F_1 = - \frac{i}{32 \pi^3} \int_0^1 \int_0^1\left[ \frac{\partial}{\partial\bar{u}} \wp(u_1 + \tau u_2,\tau)\right] \left[\frac{\partial}{\partial\bar{u}} H_1(u_1, u_2; \tau, \bar \tau)\right] \d u_1 \d u_2 .
\end{equation*}
But $\frac{\partial}{\partial\bar{u}} (u_1 + \tau u_2)=0$ and  
\begin{equation*}
\frac{\partial}{\partial\bar{u}} \frac{1}{(u_1 + \tau u_2)^2} = - \frac{\pi}{\tau_2} \frac{\partial}{\partial{u}}\left[\delta(u_1)\delta(u_2)\right]
\end{equation*}
with  $\frac{\partial}{\partial{u}} = \frac{\bar \tau \partial_{u_1} - \partial_{u_2}}{\bar \tau - \tau}$.
As the Weierstrass function has a double pole at the origin, with coefficient $1$ and no residue, we have 
\begin{equation*}
\partial_{\bar \tau} \hat F_1 = - \frac{i}{32 \pi^2 \tau_2} \left[ \frac{\partial}{\partial{u}} \frac{\partial}{\partial\bar{u}} H_1(u_1, u_2; \tau, \bar \tau) \right] \big|_{u_1 = u_2 = 0}
\end{equation*}
which gives (\ref{F1-holomorphicanomaly}) directly.

\subsection{General $k$} \label{subsec.S3k}

Generalizing (\ref{eqn.barG1}),
the supercurrent for the $\cN=(0,1)$ WZW model with bosonic level $\kappa=k-1$ is:
\begin{equation}
\bar G = \sqrt{-1} \sqrt{\frac{2}{\kappa+2}} \bar \psi_1 \bar \psi_2 \bar \psi_3 + \cdots
\end{equation}
The ellipsis is proportional to $\sum_a \psi_a J^a_b$, with $J^a_b$ being the currents of the bosonic WZW  model,
and so it cannot soak the three fermion zeromodes in the torus one-point function. 
It follows that the one-point function of $-\psi_1 \psi_2 \psi_3 \bar G$ equals 
\begin{equation}\label{eqn.gkappa}
g_\kappa(\tau, \bar \tau) = \sqrt{\frac{2}{\kappa+2}} |\eta(\tau)|^6 Z^{\mathrm{WZW}}_\kappa(\tau, \bar \tau)
\end{equation}
where 
\begin{equation} \label{eqn.bosonicWZW}
Z_\kappa^{\mathrm{WZW}}(\tau, \bar \tau) = \sum_{2j+1=1}^{\kappa+1} |\chi_j^{(\kappa)}(\tau)|^2
\end{equation}
is the torus partition function of the bosonic WZW  model, which can be expanded in terms of characters $\chi_j^{(\kappa)}(\tau)$ of 
the WZW current algebra, as the WZW CFT is rational.

 \subsubsection{Characters and source}
The characters of $\mathrm{SU}(2)_\kappa$ appearing in (\ref{eqn.bosonicWZW}) can be written as 
\begin{equation}\label{eqn.SU2char}
\chi_j^{(\kappa)}(\tau) = \frac{ \sum_{m \in Z+ \frac{j+\frac12}{\kappa+2}} q^{(\kappa+2)m^2 } \left[ 2 m(\kappa+2) \right]}{ \prod_{n>0} (1-q^n)^3}.
\end{equation}
In terms of the traditional definition of weight $3/2$ theta functions
\begin{equation*}
\Theta_{k,\ell}(\tau) = \sum_{m\in Z + \frac{\ell}{2k}} m q^{k m^2},
\end{equation*}
equation~(\ref{eqn.SU2char}) is equivalent to
\begin{equation*}
\chi_j^{(\kappa)}(\tau) = \frac{ 2(\kappa+2)\Theta_{\kappa+2,2j+1}(\tau)}{\eta(\tau)^3},
\end{equation*}
and so (\ref{eqn.gkappa}) becomes:
\begin{equation}
g_\kappa(\tau, \bar \tau) = 4 (\kappa+2) \sqrt{2(\kappa+2)}  \sum_{2j+1=1}^{\kappa+1} |\Theta_{\kappa+2,2j+1}(\tau)|^2.
\end{equation}

In \S\ref{flavoured-nullhomotopic} we will use the flavoured WZW characters
\begin{equation}
\chi_j^{(\kappa)}(\xi;\tau) = \frac{\vartheta_{\kappa+2,2j+1}(\xi;\tau)- \vartheta_{\kappa+2,2j+1}(-\xi;\tau)}{\theta(\xi;\tau)}.
\end{equation}
where \begin{equation*}
\vartheta_{k,\ell}(\xi;\tau) = \sum_{m\in Z + \frac{\ell}{2k}} x^{k m} q^{k m^2}.
\end{equation*}
It is useful to extend the definition of $\chi_j^{(\kappa)}(\xi;\tau)$ to the full range $0 \leq 2j+1 <2(\kappa+2)$,
with $\chi_{-\frac12}^{(\kappa)}(\xi;\tau) =0$ and $\chi_{\frac{\kappa+1}{2}}^{(\kappa)}(\xi;\tau) =0$.
Using 
\begin{equation*}
\vartheta_{k,\ell}(-\xi;\tau) =\vartheta_{k,2 k-\ell}(\xi;\tau)
\end{equation*}
one has 
\begin{equation*}
\chi_{\kappa+1-j}^{(\kappa)}(\xi;\tau) = \chi_j^{(\kappa)}(\xi;\tau).
\end{equation*}

\subsubsection{Solution of the holomorphic anomaly equation}
There are generalizations $\hat F_k$ of $\hat F_1$ which were described in detail in \cite{Harvey:2014cva}
and such that $2 F_{\kappa+1}$ precisely solves the correctly normalized holomorphic anomaly equation with source $g_\kappa(\tau, \bar \tau)$.

A simple way to arrive at a definition of the $\hat F_k$ is to look for an integral formula analogous to (\ref{F1-integralformula}), i.e.\ of the form:
\begin{equation}\label{Fk-integralformula}
\hat F_k = \frac{1}{8 \pi^2} \int_0^1 \int_0^1 \wp(u_1 + \tau u_2,\tau) H_k(u_1, u_2; \tau, \bar \tau) \d u_1 \d u_2 .
\end{equation}
Observe that the function 
\begin{equation}\label{formulaforHk}
H_k(u_1, u_2; \tau, \bar \tau) \equiv  \sqrt{(\kappa+2)\tau_2} e^{- (\kappa +2) \pi \tau_2 u_2^2}  \sum_{2j+1=1}^{\kappa+1} |\chi_j^{(\kappa)}(u_1 + \tau u_2;\tau)|^2 |\theta(u_1 + \tau u_2,\tau)|^2
\end{equation}
satisfies a heat equation 
\begin{equation*}
\partial_{\bar \tau} H_k(u_1, u_2; \tau, \bar \tau)= \frac{i}{2 \pi (\kappa+2)} \partial_{\bar u}^2 H_k(u_1, u_2; \tau, \bar \tau)
\end{equation*}
and furthermore $\partial_u \chi_j^{(\kappa)}(u;\tau)\theta(u,\tau)|_{u=0} = 2 \pi i \chi_j^{(\kappa)}(\tau)$. 

That means the same manipulations as in \S\ref{subsec.S31} give us 
\begin{equation*}
\partial_{\bar \tau} \hat F_k = - \frac{i}{16 \pi^2 (\kappa+2) \tau_2} \left[ \frac{\partial}{\partial{u}} \frac{\partial}{\partial\bar{u}} H_k(u_1, u_2; \tau, \bar \tau) \right] \big|_{u_1 = u_2 = 0}
\end{equation*}
which gives directly the desired
\begin{equation}
\sqrt{-2\tau_2}\partial_{\bar \tau} \hat F_k = \frac14 \sqrt{\frac{2}{\kappa+2}}\sum_{2j+1=1}^{\kappa+1} |\chi_j^{(\kappa)}(\tau)|^2.
\end{equation}

\subsubsection{Holomorphic part}

In order to extract the holomorphic part, it is useful to Poisson resum $H_k$ from (\ref{formulaforHk}).
We can split $H_k$ in two parts:
\begin{align*}
H^{(1)}_k(u_1, u_2; \tau, \bar \tau) & \equiv  \sqrt{(\kappa+2)\tau_2} e^{- (\kappa +2) \pi \tau_2 u_2^2}  \sum_{2j+1=0}^{2\kappa+3}  |\vartheta_{\kappa+2,2j+1}(\xi;\tau)|^2,
\\
H^{(2)}_k(u_1, u_2; \tau, \bar \tau) & \equiv  - \sqrt{(\kappa+2)\tau_2} e^{- (\kappa +2) \pi \tau_2 u_2^2} \sum_{2j+1=0}^{2\kappa+3}\vartheta_{\kappa+2,2j+1}(\xi;\tau)\overline{\vartheta_{\kappa+2,2 \kappa + 4-2j-1}(\xi;\tau)}.
\end{align*}
These Poisson resum to
\begin{align*}
H^{(1)}_k(u_1, u_2; \tau, \bar \tau) &= \sum_{n,m \in \bZ} (\kappa+2) e^{2 \pi i (\kappa+2)(n u_2 - m u_1)}  e^{- \frac{(\kappa+2)\pi}{\tau_2}|m \tau + n|^2},
\\
H^{(2)}_k(u_1, u_2; \tau, \bar \tau) &= - \sum_{n,m \in \bZ} e^{2 \pi i(n u_2 - m u_1)}  e^{- \frac{\pi}{(\kappa+2)\tau_2}|m \tau + n|^2},
\end{align*}
which recombine to
\begin{equation}\label{Hk-resummed}
H_k(u_1, u_2; \tau, \bar \tau) = \sum_{n,m \in \bZ} \epsilon_{n,m}^{\kappa+2} e^{2 \pi i(n u_2 - m u_1)}  e^{- \frac{\pi}{(\kappa+2)\tau_2}|m \tau + n|^2},
\end{equation}
where $\epsilon_{n,m}^{\kappa+2} =\kappa+1$ if both $n$ and $m$ are divisible by $\kappa+2$, and $ \epsilon_{n,m}^{\kappa+2} =-1$ otherwise. 

In the $\bar \tau \to - i \infty$ limit (\ref{Hk-resummed}) simplifies to 
\begin{equation*}
H_k(u_1, u_2; \tau,  - i \infty) = \sum_{n,m \in \bZ} \epsilon_{n,m}^{\kappa+2} e^{2 \pi i(n u_2 - m u_1)} e^{\frac{2 i \pi}{\kappa+2}m (m \tau + n)}
\end{equation*}
and thus 
\begin{equation}
F_k = \frac{1}{8 \pi^2} \int_0^1 \int_0^1 \wp(u_1 + \tau u_2,\tau)  \sum_{n,m \in \bZ} \epsilon_{n,m}^{\kappa+2} e^{2 \pi i(n u_2 - m u_1)} e^{\frac{2 i \pi}{\kappa+2}m (m \tau + n)}  \d u_1 \d u_2 ,
\end{equation}
We can again split  into two parts:
\begin{align}
F^{(1)}_k & = \frac{\kappa+2}{8 \pi^2} \int_0^1 \int_0^1 \wp(u_1 + \tau u_2,\tau)  \sum_{n,m \in \bZ}  e^{2 \pi i(\kappa+2)(n u_2 - m u_1)} e^{2 i \pi(\kappa+2) m^2\tau}  \d u_1 \d u_2 , \label{Fk(1)}
\\
F^{(2)}_k & = - \frac{1}{8 \pi^2} \int_0^1 \int_0^1 \wp(u_1 + \tau u_2,\tau)  \sum_{n,m \in \bZ}  e^{2 \pi i(n u_2 - m u_1)} e^{\frac{2 i \pi}{\kappa+2}m^2 \tau} e^{\frac{2 i \pi}{\kappa+2}m n}  \d u_1 \d u_2 . \label{Fk(2)}
\end{align}

The sum over $n$ in (\ref{Fk(1)}) gives a sum of delta functions at $u_2 = \frac{\ell}{\kappa+2}$, so that 
\begin{equation}
F^{(1)}_k = \sum_{\ell=0}^{\kappa+1}  \sum_{m \in \bZ} q^{(\kappa+2) m^2+ m \ell}\frac{1}{8 \pi^2}   \int_0^1 e^{-2 \pi i(\kappa+2) m u_1} \wp(u_1,\tau) \d u_1, \label{Fk(1)-int}
\end{equation}
while the sum over $n$ in (\ref{Fk(2)}) gives a delta function at $u_2 = -\frac{m}{\kappa+2}$ modulo $1$:
\begin{align}
F^{(2)}_k & = -   \sum_{m \in \bZ}  q^{\frac{m^2}{\kappa+2} +m [-\frac{m}{\kappa+2}]} \frac{1}{8 \pi^2} \int_0^1 e^{-2 \pi i m u_1}   \wp(u_1,\tau) \d u_1 , \notag
\\
& = - \sum_{\ell=0}^{\kappa+1}  \sum_{m \in \bZ}  q^{(\kappa+2) m^2 + m \ell} \frac{1}{8 \pi^2} \int_0^1 e^{-2 \pi i ((\kappa+2) m+\ell) u_1}   \wp(u_1,\tau) \d u_1 . \label{Fk(2)-int}
\end{align}
In particular, the $m=0$ contributions add up to $k G_2$, while the other ones give positive powers of $q$ with integral coefficients. 
We conclude that the invariant for the general $k$ case is $-\frac{k}{12} \pmod 2$, as expected!

\subsection{An antiholomorphic SCFT of degree $27$}

Following \cite[Chapter 13]{MR3223024}, write $\alpha \in \pi_3\mathrm{TMF}$ and $\beta \in \pi_{10}\mathrm{TMF}$ for the TMF classes  represented, respectively, by $\mathrm{SU}(2)$ and $\mathrm{Sp}(2)$ with their Lie group framings. These are the classes that, conjecturally, correspond to the antiholomorphic free-fermion SCFTs described in Puzzles~\ref{mainpuzzle} and~\ref{sp2puzzle}; the motivation for those puzzles came from the fact that $\alpha$ and $\beta$ have exact orders $24$ and $3$, respectively.
There is no TMF class with Witten genus $\Delta$, but $24\Delta$ is the Witten genus of an element in $\pi_{12}\mathrm{TMF}$; that element is, not surprisingly, called simply $\{24\Delta\}$. As observed in \cite{GJFIII}, this class can also be represented by a purely antiholomorphic SCFT. Specifically, \cite{MR2352133} constructs a holomorphic $\cN=1$ SCFT called $V^{s\natural}$ (the ``$s$'' stands for ``super'' and the ``$\natural$'' stands for ``moonshine''), and $\{24\Delta\}$ is represented by a right-moving copy $\bar{V}^{s\natural}$. 

It is known that 
$$ \alpha \times \{24\Delta\} = 0 \in \pi_{27}\mathrm{TMF}.$$
This raises already an interesting puzzle:
\begin{puzzle}
  Find a nullhomotopy for the degree-27 SQFT $\overline{\Fer}(3) \otimes \bar{V}^{s\natural}$.
\end{puzzle}

Although $\alpha \times \{24\Delta\} = 0$, there is an interesting order-$3$ class in $\pi_{27}\mathrm{TMF}$, which could be called $\{8\alpha\Delta\}$.%
\footnote{Compare \cite[Chapter 13, \S1]{MR3223024}, but note that what is reported there are the localizations on $\pi_*\mathrm{TMF}$ at the primes $2$ and $3$; the interesting $3$-local class is called simply $\{\alpha\Delta\}$, since $8$ is invertible $3$-locally.}
We will describe an antiholomorphic SCFT which we expect to represent this class. 
The supersymmetry-preserving automorphism group of $\bar{V}^{s\natural}$ is Conway's largest simply group $\mathrm{Co}_1$, and the Ramond-sector ground states for $\bar{V}^{s\natural}$ form the $24$-dimensional ``Leech lattice'' representation of the double cover $\mathrm{Co}_0 = 2.\mathrm{Co}_1$ \cite{MR2352133}. The conjugacy classes of order $3$ in $\mathrm{Co}_1$ and $\mathrm{Co}_0$ are in bijection; there are four of them, distinguished by their trace on the $24$-dimensional representation. The class called ``$3\mathrm{A}$'' in \cite{ATLAS} acts with trace $0$. Now consider the order-$3$ automorphism of $\overline{\Fer}(3) \otimes \bar{V}^{s\natural}$ which acts on $\bar{V}^{s\natural}$ by class $3\mathrm{A}$, and which acts on $\overline{\Fer}(3)$ by cyclicly permuting the three fermions. It is not hard (c.f.\ \cite{JFT,GJFIII}) to calculate that the 't Hooft anomalies on the $\overline{\Fer}(3)$ and $\bar{V}^{s\natural}$ factors cancel, and the corresponding orbifold $(\overline{\Fer}(3) \otimes \bar{V}^{s\natural}) \sslash \bZ_3$ is our proposed representative of $\{8\alpha\Delta\}$.

Indeed, the one-point function of $\bar{G}$ in this orbifold is $8|\eta|^6\Delta$, and so the calculations in \S\ref{subsec.S31} give the solution $16  \hat{F}_1\Delta$ to the holomorphic anomaly equation. The holomorphic part of this solution is
$$ 16  F_1\Delta = -\frac23 q + q^2 2\bZ[\![q]\!], $$
showing that the SQFT $(\overline{\Fer}(3) \otimes \bar{V}^{s\natural}) \sslash \bZ_3$ has order (at least) $3$.

\subsection{Is $S^3_k$ flavoured-nullhomotopic?}\label{flavoured-nullhomotopic}

The flavoured elliptic genus for the $(0,1)$ $\mathrm{SU}(2)_\kappa$ WZW model, with flavour fugacity for the chiral $\mathrm{SU}(2)$ rotations, 
takes the form (after some reorganization)
\begin{equation}
\sqrt{2(\kappa+2)} \sum_{2j+1=0}^{2\kappa+3}\frac{\vartheta_{\kappa+2,2j+1}(2\xi;\tau)}{\theta(2\xi;\tau)}\Theta_{\kappa+2,2j+1}(\bar \tau).
\end{equation}
It turns out that there are interesting solutions of the holomorphic anomaly equation with this source, which 
have {\it integral} coefficients, but are meromorphic in $\xi$ with a double pole at $\xi=0$. 

These are built from the modular completions of the Appell--Lerche sums
\begin{equation}
{\cal A}_{1,\kappa+2}(\xi;\tau) = \sum_{m\in Z} q^{(\kappa+2)m^2} x^{2 m (\kappa+2)}\frac{1+ x q^m}{1-x q^m}.
\end{equation}
The precise statement is that 
${\cal A}_{1,\kappa+2}(\xi;\tau)$ can be completed to a Jacobi object $\hat {\cal A}_{1,\kappa+2}(\xi;\tau)$ by combining it with a multiple of 
\begin{equation*}
\sum_{2j+1=0}^{2\kappa+3}\vartheta_{\kappa+2,2j+1}(\xi;\tau)\Theta^*_{\kappa+2,2j+1}(\bar \tau)
\end{equation*}
where 
\begin{equation*}
\Theta^*_{k,\ell}(\tau) = \frac12 \sum_{m\in Z + \frac{\ell}{2k}} \mathrm{sign}(m) \mathrm{erfc}(2 |m| \sqrt{\pi k \tau_2}) q^{-k m^2},
\end{equation*}
which satisfies the holomorphic anomaly equation with source 
\begin{equation*}
\sqrt{2(\kappa+2)} \sum_{2j+1=0}^{2\kappa+3}\vartheta_{\kappa+2,2j+1}(2\xi;\tau)\Theta_{\kappa+2,2j+1}(\bar \tau).
\end{equation*}

It follows that 
\begin{equation}\label{eqn.Jk}
J_k \equiv \frac{\eta(\tau)^3}{\theta(2\xi;\tau)} {\cal A}_{1,\kappa+2}(\xi;\tau)
\end{equation}
is the holomorphic part of a solution to  the holomorphic anomaly equation sourced by the flavoured $S^3_k$ torus one-point function. 
It has a double pole as $\xi \to 0$. Its $q$-expansion has integral coefficients. 

In particular, (\ref{eqn.Jk}) is a natural candidate for the flavoured elliptic genus for some SQFT which has $S^3_k$ as one boundary component and 
some other boundary component with a geometry analogous to $\bR^2 \times \bR^2$, with both planes rotated by the flavour symmetry, so as to be flavoured-compact in the sense of \S\ref{subsec.flavoured-compact}. 
\begin{puzzle}Identify the corresponding $(0,4)$ four-dimensional ``trumpet'' geometry. \end{puzzle}

In principle, given a group $G$, one may define a version of ``$G$-flavoured topological modular forms'' by working with the derived moduli stack of elliptic curves equipped with sufficiently-nondegenerate $G$-bundle. Such a theory should have classes represented by flavoured-compact manifolds, and this ``trumpet'' should be a ``flavoured nullcobordism'' of $S^3_k$.

\acknowledgments\addcontentsline{toc}{section}{Acknowledgments}
Research at the Perimeter Institute is supported by the Government of Canada through Industry Canada and by the Province of Ontario through the Ministry of Economic Development and Innovation.
We thank S.\ Murthy and E.\ Witten for discussions.


\newcommand{\etalchar}[1]{$^{#1}$}

\end{document}